\documentclass[12pt, reqno]{amsart}

\usepackage[margin=.9in, a4paper, foot=.3in]{geometry}

\usepackage{graphicx, color}

\definecolor{darkblue}{rgb}{0,0,.5}
\definecolor{darkred}{rgb}{.5,0,0}
\definecolor{darkgreen}{rgb}{0,0.5,0}

\usepackage{hyperref}
\hypersetup {
    pdfstartview=FitH,
    colorlinks,
    citecolor=darkgreen,
    linkcolor=darkred,
    urlcolor=darkblue
}

\let\oldtocsection=\tocsection
\let\oldtocsubsection=\tocsubsection
\let\oldtocsubsubsection=\tocsubsubsection

\renewcommand{\tocsection}[2]{\hspace{0em}\oldtocsection{#1}{#2}}
\renewcommand{\tocsubsection}[2]{\hspace{1em}\oldtocsubsection{#1}{#2}}
\renewcommand{\tocsubsubsection}[2]{\hspace{2em}\oldtocsubsubsection{#1}{#2}}

\usepackage{cite}

\usepackage[noBBpl,slantedGreek]{mathpazo}
\usepackage[mathcal]{euscript}

\usepackage{amssymb}
\allowdisplaybreaks
\numberwithin{equation}{section}
\numberwithin{figure}{section}

\usepackage{bbm}

\usepackage{mathtools}

\usepackage{ifthen}

\newcommand {\svee}[2][\null]{#2%
  \ifthenelse{\equal{#1}{\null}}{\,}{}%
  \check{{}_{#1}}%
  \ifthenelse{\equal{#1}{\null}}{\,}{}%
}

\newcommand {\interval}[2]{[#1 \, . \, . \, #2]}

\newcommand {\id}{\mathrm{id}}
\newcommand {\rmd}{\mathrm d}
\newcommand {\rme}{\mathrm e}

\newcommand {\bbC}{\mathbb C}
\newcommand {\bbF}{\mathbb F}
\newcommand {\bbZ}{\mathbb Z}

\newcommand {\calM}{\mathcal M}
\newcommand {\calR}{\mathcal R}
\newcommand {\calT}{\mathcal T}
\newcommand {\calV}{\mathcal V}

\newcommand {\gothh}{\mathfrak h}
\newcommand {\gothg}{\mathfrak g}
\newcommand {\tgothh}{\widetilde{\mathfrak h}}
\newcommand {\uqlg}{\mathrm U_q(\mathcal L(\mathfrak g))}
\newcommand {\uqlslii}{\mathrm U_q(\mathcal L(\mathfrak{sl}_2))}
\newcommand {\uqlsliii}{\mathrm U_q(\mathcal L(\mathfrak{sl}_3))}
\newcommand {\uqlsllpo}{\mathrm U_q(\mathcal L(\mathfrak{sl}_{l + 1}))}

\DeclareMathOperator {\diag}{diag}
\DeclareMathOperator {\End}{End}
\DeclareMathOperator {\tr}{tr}

\title{Reduced qKZ equation: general case}

\author[A. Kl\"umper]{Andreas Kl\"umper}
\address{Mathematics and Natural Sciences, University of Wuppertal, 42097 Wuppertal, Germany}
\email{kluemper@uni-wuppertal.de}

\author[Kh. S. Nirov]{\vskip .2em Khazret S. Nirov}
\address{Institute for Nuclear Research of the Russian Academy of Sciences, 60th October Ave 7a, 117312 Moscow, Russia}
\address{Faculty of Mathematics, National Research University ``Higher School of Economics'', 119048 Moscow, Russia}
\curraddr{Mathematics and Natural Sciences, University of Wuppertal, 42097 Wuppertal, Germany}
\email{nirov@uni-wuppertal.de}

\author[A. V. Razumov]{Alexander V. Razumov}
\address{Institute for High Energy Physics, NRC ``Kurchatov Institute", 142281 Protvino, Mos\-cow region, Russia}
\email{Alexander.Razumov@ihep.ru}

\begin{document}

\begin{abstract}
We use the quantum group approach for the investigation of correlation
functions of integrable vertex models and spin chains. For the inhomogeneous
reduced density matrix in case of an arbitrary simple Lie algebra we find
functional equations of the form of the reduced quantum Knizhnik-Zamolodchikov
equation. This equation is the starting point for the investigation of correlation
functions at arbitrary temperature and notably for the ground state.
\end{abstract}

\maketitle

\tableofcontents

\section{Introduction}

In this paper we derive a difference-type functional equation, called the discrete reduced quantum Knizhnik--Zamolodchikov equation, for the density operator of a quantum integrable vertex model related to an arbitrary complex simple Lie algebra.
Our setting allows for the study of correlation functions at finite and zero temperature in the thermodynamic limit, or alternatively of ground-state correlators on finite ring shaped and infinite chains. Throughout this paper we use methods based on the notion of a quantum group introduced by Drinfeld \cite{Dri87} and Jimbo \cite{Jim85}. To be precise, we consider quantum integrable systems related to a special class of quantum groups, namely the quantum loop algebras, see section \ref{ss:qla} for the definition.

Our work aims at extending the previous work, see \cite{JimMikMiwNak92, JimMiw96} and later developments, related to systems based on the quantum loop algebra $\uqlslii$ which enjoys a simple crossing symmetry due to the equivalence of any representation with its dual. Some explorative investigations for a system based on the first fundamental representation of $\uqlsliii$ allowed for the computation of nearest and next-nearest neighbour correlators for the associated quantum spin chain of XXX-type in the ground-state \cite{BooHutNir18, RibKlu19}. In these works the necessity of dealing simultaneously with at least two different representations of the same quantum group became obvious. Furthermore, unitarity conditions involving different
representations, and crossing relations for representations dual to each other
appeared. Here we put such constructions on solid systematic grounds valid for
arbitrary representations of any quantum group. Our constructions will allow
for a uniformized investigation of correlation functions making ad-hoc
constructions obsolete.

The central object of the quantum group approach is the universal $R$-matrix being an element of the tensor product of two copies of the quantum loop algebra. The integrability objects are constructed by choosing representations for the factors of that tensor product.\footnote{For the corresponding terminology we refer to our paper \cite{BooGoeKluNirRaz14a}.} The consistent application of the method for constructing integrability objects and proving their properties was initiated by Bazhanov, Lukyanov and Zamolodchikov \cite{BazLukZam96, BazLukZam97, BazLukZam99}. They studied the quantum version of KdV theory. Later on the method proved to be efficient for studying other quantum integrable models. Accordingly, within the framework of this approach, $R$-operators \cite{KhoTol92, LevSoiStu93, ZhaGou94, BraGouZhaDel94, BraGouZha95, BooGoeKluNirRaz10, BooGoeKluNirRaz11},
monodromy operators and $L$-operators were constructed \cite{BazTsu08,  BooGoeKluNirRaz10, BooGoeKluNirRaz11, BooGoeKluNirRaz13, Raz13, BooGoeKluNirRaz14a}. The corresponding sets of functional relations were found and proved \cite{BazHibKho02, Koj08, BazTsu08, BooGoeKluNirRaz14a, BooGoeKluNirRaz14b, NirRaz14}.

To derive the reduced qKZ equation one needs some special properties of the
integrability object related to the quantum loop algebra under consideration. Namely, one uses the unitarity relations, crossing relations and the so-called initial condition. It appears that these relations, apart from the initial condition, follow from the properties of the universal $R$-matrix. The detailed discussion can be found in paper \cite{NirRaz18}, see also paper \cite{FreRes92}. 

The plan of the paper is as follows. In section \ref{s:qla} we introduce a quantum loop algebra, its universal $R$-matrix, and define the basic integrability objects
called $R$-operators. Then we describe the properties of $R$-operators, such as the unitarity and crossing relations, necessary for the subsequent derivation of the reduced qKZ equation. This section is concluded by the definition of monodromy and transfer operators.

In section \ref{s:do} we discuss the construction of the Hamiltonian of the
system as a member of the system of commuting quantities. The aforementioned initial condition is also given here. We introduce a convenient normalization of the $R$-operators which leads to a simple form of the crossing and unitarity relations. Also the initial condition becomes simple. Then we remind of the definition of the density operator and represent it as the Trotter limit of some sequence of operators. Such a representation allows us to relate the density operator to the partition sum of some square lattice vertex model with the free horizontal boundaries.

A graphical derivation of the reduced qKZ equation is described in section \ref{s:rqe}, and the corresponding pictures with appropriate comments are placed in the appendix.

\section{Quantum loop algebras and integrability objects} \label{s:qla}

\subsection{Preliminaries on Lie algebras}

Let $\gothg$ be a complex finite dimensional simple Lie algebra of rank $l$ \cite{Ser01, Hum80}, $\gothh$ a Cartan subalgebra of $\gothg$, and $\Delta$ the root system of $\gothg$ relative to $\gothh$. Fix a system of simple roots $\alpha_i$, $i \in \interval{1}{l}$. It is known that the corresponding coroots $h_i$ form a basis of $\gothh$, so that
\begin{equation*}
\gothh = \bigoplus_{i = 1}^l \bbC \, h_i.
\end{equation*}
The Cartan matrix $A = (a_{i j})_{i, \, j \in \interval{1}{l}}$ of $\gothg$ is defined by the equation
\begin{equation*}
a_{i j} = \langle \alpha_j, \, h_i \rangle.
\end{equation*}

Denote by $\theta$ the highest root of $\gothg$ \cite{Ser01, Hum80}. We have
\begin{equation*}
\theta = \sum_{i = 1}^l a_i \, \alpha_i, \qquad \svee{\theta} = \sum_{i = 1}^l \svee[i]{a} \, h_i
\end{equation*}
for some positive integers $a_i$ and $\svee[i]{a}$ with $i \in \interval{1}{l}$. These integers, together with
\begin{equation*}
a_0 = 1, \qquad \svee[0]{a} = 1,
\end{equation*}
are the Kac labels and the dual Kac labels of the Dynkin diagram associated with the extended Cartan matrix $A^{(1)}$. Recall that the sums
\begin{equation*}
h = \sum_{i = 0}^l a_i, \qquad \svee{h} = \sum_{i = 0}^l \svee[i]{a}
\end{equation*}
are called the Coxeter number and the dual Coxeter number of $\gothg$.

Denote by $\tgothh$ the Cartan subalgebra of $\gothg$ extended by a one dimensional center $\bbC \, K$. We consider the simple roots  $\alpha_i$, $i \in \interval{1}{l}$, as elements of $\tgothh^*$ assuming that
\begin{equation*}
\langle \alpha_i, \, K \rangle = 0.
\end{equation*}
Introduce an additional `root'
\begin{equation*}
\alpha_0 = - \theta
\end{equation*}
and an additional `coroot'
\begin{equation*}
h_0 = K - \svee{\theta}.
\end{equation*}
After that for the entries of the extended Cartan matrix $A^{(1)} = (a_{i j})_{i, \, j \in \interval{0}{l}}$ of $\gothg$ we have the expression
\begin{equation*}
a_{i j} = \langle \alpha_j, \, h_i \rangle.
\end{equation*}

\subsection{Quantum loop algebras} \label{ss:qla}

Let $\hbar$ be a nonzero complex number such that $q = \exp \hbar$ is not a root of unity. We assume that
\begin{equation*}
q^\nu = \exp (\hbar \nu)
\end{equation*}
for any $\nu \in \bbC$. As usually, we define the $q$-deformation of a number $\nu \in \bbC$ as
\begin{equation*}
[\nu]_q = \frac{q^\nu - q^{-\nu}}{q - q^{-1}}.
\end{equation*}

Note that the extended Cartan matrix $A^{(1)}$ is symmetrizable. It means that there exists a diagonal matrix $D = \diag(d_0, \, d_1, \, \ldots, d_l)$, where $d_i$, $i \in \interval{0}{l}$, are positive integers, such that the matrix $D A^{(1)}$ is symmetric. Such a matrix $D$ is defined up to a nonzero scalar factor. We fix the integers $d_i$ assuming that they are relatively prime and denote
\begin{equation*}
q_i = q^{d_i}. 
\end{equation*}

The quantum loop algebra $\uqlg$ is a unital associative $\bbC$-algebra generated by the elements
\begin{equation*}
e_i, \quad f_i, \quad i = 0, 1,\ldots,l, \qquad q^x, \quad x \in \tgothh,
\end{equation*}
satisfying the relations
\begin{gather}
q^{\nu K} = 1, \quad \nu \in \bbC, \qquad q^{x_1} q^{x_2} = q^{x_1 + x_2}, \label{djra} \\
q^x e_i \, q^{-x} = q^{\langle \alpha_i, \, x \rangle} e_i, \qquad q^x f_i \, q^{-x} = q^{- \langle \alpha_i, \, x \rangle} f_i, \label{djrb} \\
[e_i, \, f_j] = \delta_{i j} \, \frac{q_i^{h_i} - q_i^{- h_i}}{q^{\mathstrut}_i - q_i^{-1}}, \label{djrc} \\
\ \sum_{n = 0}^{1 - a_{i j}} (-1)^n \frac{e_i^{1 - a_{i j} - n}}{[1 - a_{i j} - n]_{q^i}!} e^{\mathstrut}_j \, \frac{e_i^n}{[n]_{q^i}!} = 0, \qquad \sum_{n = 0}^{1 - a_{i j}} (-1)^n \frac{f_i^{1 - a_{i j} - n}}{[1 - a_{i j} - n]_{q^i}!} f^{\mathstrut}_j \, \frac{f_i^n}{[n]_{q^i}!} = 0. \label{djrd}
\end{gather}
Here, relations (\ref{djrb}) and (\ref{djrc}) are valid for all $i, j \in \interval{0}{l}$. 
The last line of the relations is valid for all distinct $i, j \in \interval{0}{l}$.

The quantum loop algebra $\uqlg$ is a Hopf algebra. Here the multiplication mapping 
$\mu \colon \uqlg \otimes \uqlg \to \uqlg$ is defined as
\begin{equation*}
\mu(a \otimes b) = a b,
\end{equation*}
and for the unit mapping $\iota \colon \bbC \to \uqlg$ we have
\begin{equation*}
\iota(\nu) = \nu \, 1.
\end{equation*} 
The comultiplication $\Delta$, the antipode $S$, and the counit $\varepsilon$ are given by the relations
\begin{gather}
\Delta(q^x) = q^x \otimes q^x, \qquad \Delta(e^{}_i) = e^{}_i \otimes 1 + q_i^{h_i} \otimes e^{}_i, \qquad \Delta(f^{}_i) = f^{}_i \otimes q_i^{- h_i} + 1 \otimes f^{}_i, \label{hsa} \\
S(q^x) = q^{- x}, \qquad S(e^{}_i) = - q_i^{- h_i} e^{}_i, \qquad S(f^{}_i) = - f^{}_i \, q_i^{h_i}, \label{sg} \\
\varepsilon(q^h) = 1, \qquad \varepsilon(e^{}_i) = 0, \qquad \varepsilon(f^{}_i) = 0. \label{hsc}
\end{gather}
For the inverse of the antipode one has
\begin{equation}
S^{-1}(q^x) = q^{-x}, \qquad S^{-1}(e^{}_i) = -e^{}_i \, q_i^{-h_i}, \qquad S^{-1}(f^{}_i) = - q_i^{h_i} f^{}_i. \label{isg}
\end{equation}

\subsection{\texorpdfstring{Universal $R$-matrix}{Universal R-matrix}}

Let $\Pi$ be the automorphism of the tensor square of the algebra $\uqlg$ defined by the equation
\begin{equation*}
\Pi (a \otimes b) = b \otimes a.
\end{equation*}
It is known that the mapping 
\begin{equation*}
\Delta' = \Pi \circ \Delta
\end{equation*}
is a comultiplication in $\uqlg$ called the opposite comultiplication.

Let $\uqlg$ be a quantum loop algebra. There exists a unique element $\calR$ of the tensor product of $\uqlg \otimes \uqlg$ connecting the two comultiplications as
\begin{equation*}
\Delta'(a) = \calR \, \Delta(a) \calR^{-1}
\end{equation*}
for any $a \in \uqlg$, and satisfying in $\uqlg \otimes \uqlg \otimes \uqlg$ the equations
\begin{equation*}
(\Delta \otimes \id)(\calR) = \calR^{(13)} \, \calR^{(23)}, \qquad (\id \otimes \Delta)(\calR) = \calR^{(13)} \, \calR^{(12)}.
\end{equation*}
The meaning of the superscripts in the above relations is explained in any textbook on quantum groups, see also the appendix of paper \cite{NirRaz18}. The element $\calR$ is called the universal $R$-matrix. One can show that it satisfies the universal Yang-Baxter equation
\begin{equation*}
\calR^{(12)} \, \calR^{(13)} \, \calR^{(23)} = \calR^{(23)} \, \calR^{(13)} \, \calR^{(12)}
\end{equation*}
in $\uqlg \otimes \uqlg \otimes \uqlg$.

There are two main approaches to the construction of the universal $R$-matrix for a quantum loop algebra. One of them was proposed by Khoroshkin and Tolstoy \cite{TolKho92, KhoTol92, KhoTol93, KhoTol94}, and another one is related to the names of Beck and Damiani \cite{Bec94a, Dam98}. It should be noted that we define the quantum loop algebra as a $\bbC$-algebra. It can be also defined as a $\bbC[[\hbar]]$-algebra, where $\hbar$ is considered as an indeterminate. In this case one really has the universal $R$-matrix. In our case, the universal $R$-matrix exists only in some restricted sense, see, for example, paper \cite{Tan92}, and the corresponding discussion in paper \cite{NirRaz18} for the case of $\uqlsllpo$.

As for any Hopf algebra, starting from two representations of $\uqlg$, say $\varphi_1$ and $\varphi_2$, we construct a new representation $\varphi_1 \otimes_\Delta \varphi_2$ of $\uqlg$ by the relation
\begin{equation*}
\varphi_1 \otimes_\Delta \varphi_2 = (\varphi_1 \otimes \varphi_2) \circ \Delta.
\end{equation*}
The corresponding $\uqlg$-module is denoted by $V_1 \otimes_\Delta V_2$, where $V_1$ and $V_2$ are the modules corresponding to the representations $\varphi_1$ and $\varphi_2$.

\subsection{Spectral parameter}

In applications to the theory of quantum integrable systems, one usually considers families of representations of a quantum loop algebra parametrized by a complex parameter called a spectral parameter. We introduce a spectral parameter in the following way. Assume that the quantum loop algebra $\uqlg$ is $\bbZ$-graded,
\begin{equation*}
\uqlg = \bigoplus_{m \in \bbZ} \uqlg_m, \qquad \uqlg_m \, \uqlg_n \subset \uqlg_{m + n},
\end{equation*}
so that any element $a \in \uqlg$ can be uniquely represented as
\begin{equation*}
a = \sum_{m \in \bbZ} a_m, \qquad a_m \in \uqlg_m.
\end{equation*}
Given $\zeta \in \bbC^\times$, we define the grading automorphism $\Gamma_\zeta$ by the equation
\begin{equation*}
\Gamma_\zeta(a) = \sum_{m \in \bbZ} \zeta^m a_m.
\end{equation*}
It is worth noting that
\begin{equation}
\Gamma_{\zeta_1 \zeta_2} = \Gamma_{\zeta_1} \circ \Gamma_{\zeta_2} \label{gzz}
\end{equation}
for any $\zeta_1, \zeta_2 \in \bbC^\times$. Now, for any representation $\varphi$ of $\uqlg$ we define the corresponding family $\varphi_\zeta$ of representations as
\begin{equation*}
\varphi_\zeta = \varphi \circ \Gamma_\zeta.
\end{equation*}
If $V$ is the $\uqlg$-module corresponding to the representation $\varphi$, we denote by $V_\zeta$ the $\uqlg$-module corresponding to the representation $\varphi_\zeta$.

The common way to endow $\uqlg$ by a $\bbZ$-gradation is to assume that
\begin{equation*}
q^x \in \uqlg_0, \qquad e_i \in \uqlg_{s_i}, \qquad f_i \in \uqlg_{-s_i},
\end{equation*}
where $s_i$ are arbitrary integers. It is clear that for such a $\bbZ$-gradation one has
\begin{equation*}
\Gamma_\zeta(q^x) = q^x, \qquad \Gamma_\zeta(e_i) = \zeta^{s_i} e_i, \qquad \Gamma_\zeta(f_i) = \zeta^{-s_i} f_i.
\end{equation*}
 We denote
\begin{equation*}
s = \sum_{i = 0}^l a_i s_i,
\end{equation*}
where, as above, $a_i$ are the Kac labels of the Dynkin diagram associated with the extended Cartan matrix of $\gothg$.

It follows from the explicit expression for the universal $R$-matrix~\cite{TolKho92, KhoTol92, KhoTol93, KhoTol94, Bec94a, Dam98} that
\begin{equation}
(\Gamma_\zeta \otimes \Gamma_\zeta)(\calR) = \calR \label{gagar}
\end{equation}
for any $\zeta \in \bbC$. Besides, equations (\ref{sg}) and (\ref{isg}) give
\begin{equation*}
S \circ \Gamma_\zeta = \Gamma_\zeta \circ S, \qquad S^{-1} \circ \Gamma_\zeta = \Gamma_\zeta \circ S^{-1}.
\end{equation*}

\subsection{\texorpdfstring{$R$-operators}{R-operators}}

Now recall the definition of an $R$-operator. Let $V$ and $W$ be $\uqlg$-modules and $\varphi$ and $\psi$ the corresponding representations of $\uqlg$.\footnote{In this paper we assume that all $\uqlg$-modules under consideration are finite dimensional.} The $R$-operator $R_{V | W}(\zeta | \eta)$ is defined as 
\begin{equation*}
\rho_{V | W}(\zeta | \eta) R_{V | W}(\zeta | \eta) = (\varphi_\zeta \otimes \psi_\eta)(\calR).
\end{equation*}
Here $\zeta$ and $\eta$ are spectral parameters, and $\rho_{V | W}(\zeta | \eta)$ the normalization factor.

Using~(\ref{gzz}) and~(\ref{gagar}), one can demonstrate that
\begin{equation*}
(\varphi_{\zeta \nu} \otimes \psi_{\eta \nu})(\calR) = (\varphi_\zeta \otimes \psi_\eta)(\calR)
\end{equation*}
for any $\nu \in \bbC^\times$. Therefore, under an appropriate choice of the normalization factor, $R_{V | W}(\zeta | \eta)$ depends only on the combination $\zeta \eta^{-1}$ and one can use $R$-operators depending on only one spectral parameter. Below we always use this choice of the normalization, however, for our purposes it is more convenient to consider $R$-operators as depending on two spectral parameters.

\begin{figure}[ht]
\centering
\begin{minipage}{0.4\textwidth}
\centering
\includegraphics{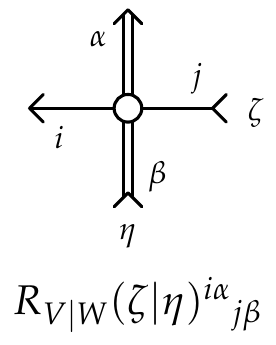}
\caption{}
\label{f:ro}
\end{minipage} \hfil
\begin{minipage}{0.4\textwidth}
\centering
\includegraphics{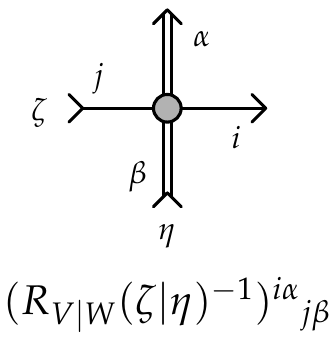}
\caption{}
\label{f:iro}
\end{minipage}
\end{figure}

We use for the matrix elements of $R_{V | W}(\zeta_1 | \zeta_2)$ the depiction which can be seen in figure~\ref{f:ro}.
Here we associate with $V$ and $W$ a single and a double lines respectively. It is worth to note that the indices in the graphical image go clockwise.

\begin{figure}[ht]
\centering
\includegraphics{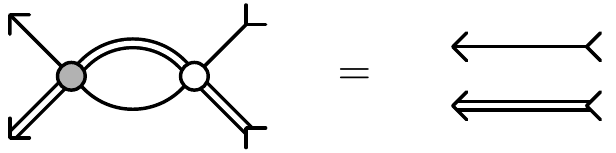}
\caption{}
\label{f:irr}
\end{figure}

For the matrix elements of the inverse $R_{V | W}(\zeta | \eta)^{-1}$ of the $R$-operator $R_{V | W}(\zeta | \eta)$ we use the depiction given in figure \ref{f:iro}. Here we use a grayed circle for the operator and the counter-clockwise 
order for the indices. This allows one to have a natural graphical form 
of the equation 
\begin{equation*}
R_{V | W}(\zeta | \eta)^{-1} R_{V | W}(\zeta | \eta) = 1_{V \otimes W},
\end{equation*}
see figure \ref{f:irr}.  

\subsection{Unitarity relations}

Let the $\uqlg$-modules $V$ and $W$ are such that the module $V_\zeta \otimes_\Delta W_\eta$ is simple for general values of the spectral parameters $\zeta$ and $\eta$. In this case the following unitarity relation
\begin{equation*}
\check R_{V | W}(\zeta | \eta) \check R_{W | V}(\eta | \zeta) = C_{V | W}(\zeta | \eta) \, 1_{W \otimes V}
\end{equation*}
is valid. Here and in similar cases below we use the notations
\begin{equation*}
\check R_{V | W}(\zeta | \eta) = P_{V | W}  R_{V | W}(\zeta | \eta), \qquad \check R_{W | V}(\eta | \zeta) = P_{W | V}  R_{W | V}(\eta | \zeta)
\end{equation*}
with $P_{V | W}$ and $P_{W | V}$ being the permutation operators on the corresponding tensor products. 

\subsection{Crossing relations}

For any finite dimensional $\uqlg$-module $V$ one has two dual modules. One
dual module is denoted by $V^*$ and is defined with the help of the antipode $S$,
another one is denoted by ${}^* \! V$ and is defined with the help of the inverse of
the antipode $S^{-1}$.
  
By a crossing relation we mean any relation connecting an $R$-operator $R_{V |
  W}(\zeta | \eta)$ with an $R$-operator for which one of the modules $V$ and
$W$ (or both) is (are) replaced by a dual module. In this paper we will use the
following three crossing relations. The first one is
\begin{equation}
R_{V^* | W}(\zeta | \eta) = \rho_{V^* | W}(\zeta | \eta)^{-1} \, \rho_{V | W}(\zeta | \eta)^{-1} \, (R_{V | W}(\zeta | \eta)^{-1})^{t_1},
 \label{ca}
\end{equation}
and the second one is
\begin{equation}
R_{V^* | W^*}(\zeta | \eta) = \rho_{V^* | W^*}(\zeta | \eta)^{-1} \, \rho_{V | W}(\zeta | \eta) \, R_{V | W}(\zeta | \eta)^t.
 \label{cb}
\end{equation}
The double dual representation $\varphi^{**}_\zeta$ is isomorphic to $\varphi_\zeta$ up to a redefinition of the spectral parameter. This leads to the third crossing relation. To describe it, we introduce the following element  
\begin{equation*}
x = - \sum_{i, j = 1}^l (2 d_i - (\theta | \theta) \svee{h} s_i/s) \, b_{i j} \, h_j
\end{equation*}
of $\tgothh$, see \cite{NirRaz18}. Here $b_{i j}$ are the matrix elements of
the matrix $B$ inverse to the Cartan matrix $A$ of the Lie algebra $\gothg$,
and $(\cdot | \cdot)$ denotes invariant nondegenerate symmetric bilinear form
on $\gothg$ normalized by the equation
\begin{equation*}
(\alpha_i | \alpha_i) = 2 d_i.
\end{equation*}
Now one can demonstrate that
\begin{multline}
(X_V \otimes 1_W) \, R_{V | W}(q^{-(\theta | \theta) \svee{h} / s} \zeta | \eta) \, (X_V^{-1} \otimes 1_W) \\ = \rho_{V | W}(q^{- (\theta | \theta) \svee{h} / s} \zeta | \eta)^{-1} \, \rho_{V^* | W}(\zeta | \eta)^{-1} \,  (R_{V^* | W^{}}(\zeta | \eta)^{-1})^{t_1}, \label{cc}
\end{multline}
where
\begin{equation*}
X_V = \varphi(q^x).
\end{equation*}
This is the third crossing relation we need. More crossing relations and the corresponding proofs can be found in paper \cite{NirRaz18}.

\subsection{Monodromy and transfer operators}

\begin{figure}[ht]
\centering
\includegraphics{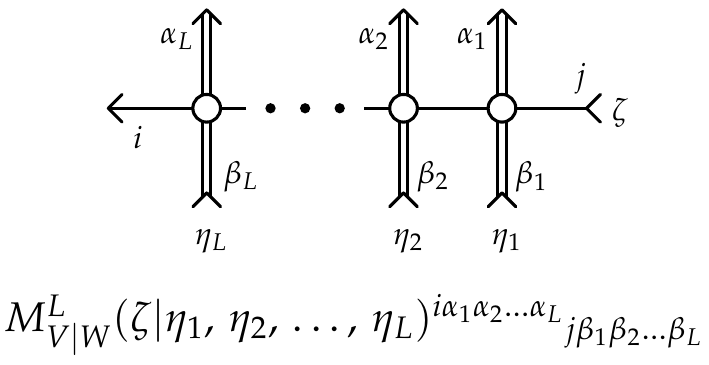}
\caption{}
\label{f:mo}
\end{figure}

In the theory of quantum integrable statistical systems the matrix elements of an $R$-operator are treated as weights of the vertices of a square lattice. To find the corresponding partition function one introduces monodromy operators and the corresponding transfer operators. To define a monodromy operator we use instead of the $\uqlg \otimes \uqlg$-module $V_\zeta \otimes W_\eta$, used in the definition of the $R$-operators, the $\uqlg \otimes \uqlg$-module
\begin{equation}
V_\zeta \otimes (W_{1 \eta_1} \otimes_\Delta W_{2 \eta_2} \otimes_\Delta \cdots \otimes_\Delta W_{L \eta_L}), \label{vw}
\end{equation}
and define the monodromy operator $M_{V | W_1, \, W_2 \, \ldots, \, W_L}(\zeta | \eta_1, \, \eta_2, \, \ldots, \, \eta_L)$ as
\begin{multline*}
\rho_{V | W_1} (\zeta | \eta_1) \, \rho_{V | W_2} (\zeta | \eta_2) \, \ldots \, \rho_{V | W_L} (\zeta | \eta_L) \, M_{V | W_1, \, W_2 \, \ldots, \, W_L}(\zeta | \eta_1, \, \eta_2, \, \ldots, \, \eta_L) \\*
= (\varphi_\zeta \otimes (\psi_{1 \eta_1} \otimes_\Delta \psi_{2 \eta_2} \otimes_\Delta \cdots \otimes_\Delta \psi_{L \eta_L}))(\calR).
\end{multline*}
Using properties of the universal $R$-matrix one can see that
\begin{equation}
M_{V | W_1, \, W_2 \, \ldots, \, W_L}(\zeta | \eta_1, \, \eta_2, \, \ldots, \, \eta_L) = R^{(1, L + 1)}_{V | W_L}(\zeta | \eta_L) \ldots R^{(1 3)}_{V | W_2}(\zeta | \eta_2) R^{(1 2)}_{V | W_1}(\zeta | \eta_1). \label{mvw}
\end{equation}
Here the meaning of the superscripts can be found again in any textbook on quantum groups. The factors of the tensor product (\ref{vw}) are numbered from left to right. The graphical representation of the matrix elements of the monodromy operator for the case $W_1 = W_2 = \cdots = W_L = W$ can be found in figure \ref{f:mo}.
The modification needed for the general case is evident.

\begin{figure}[ht]
\includegraphics{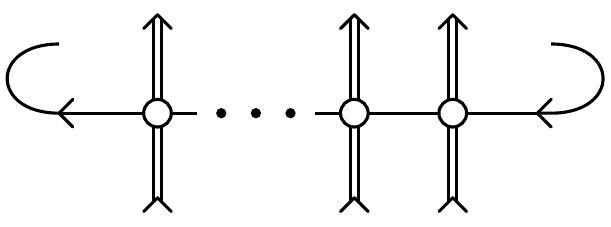}
\caption{}
\label{f:to}
\end{figure}

The transfer operator corresponding to the monodromy operator (\ref{mvw}) is defined by the equation
\begin{equation}
T_{V | W_1, \, W_2 \, \ldots, \, W_L}(\zeta | \eta_1, \, \eta_2, \, \ldots, \, \eta_L) = \tr_V (M_{V | W_1, \, W_2 \, \ldots, \, W_L}(\zeta | \eta_1, \, \eta_2, \, \ldots, \, \eta_L) \label{tvw}
\end{equation}
with the depiction for the case $W_1 = W_2 = \cdots = W_L = W$ given in figure~\ref{f:to}.
Here $\tr_V$ means the partial trace with respect to the space $V$, see, for example, the appendix of paper \cite{NirRaz18}, and hooks at the ends of the line mean that it is closed in an evident way. The most important property of transfer operators is their commutativity
\begin{equation}
[T_{V_1 | W_1, \, W_2 \, \ldots, \, W_L}(\zeta_1 | \eta_1, \, \eta_2, \, \ldots, \, \eta_L), \, T_{V_2 | W_1, \, W_2 \, \ldots, \, W_L}(\zeta_2 | \eta_1, \, \eta_2, \, \ldots, \, \eta_L)] = 0. \label{tvwtvw}
\end{equation}
It is the source of commuting quantities of quantum integrable systems.

\section{Density operator} \label{s:do}

\subsection{Commuting quantities and Hamiltonian}

The transfer operator (\ref{tvw}) acts on the $\uqlg$-module $W_{\eta_1} \otimes_\Delta \cdots \otimes_\Delta W_{\eta_L}$. As a vector space it is just $W^{\otimes L}$. We assume that $W = V$ and construct commuting quantities on $V^{\otimes L}$ as follows. First of all we denote
\begin{equation*}
T_L(\zeta) = T_{V | V}(\zeta | \underbracket[.6pt]{1, \, 1, \, \ldots, \, 1}_L).
\end{equation*}
It follows from (\ref{tvwtvw}) that the quantities
\begin{equation*}
I_m = \left. \Big(\zeta \frac{\rmd}{\rmd \zeta} \Big)^m \log T_L(\zeta) \right|_{\zeta = 1} 
\end{equation*}
commute,
\begin{equation}
[I_m, \, I_n] = 0, \qquad m, n \in \bbZ_{>0}. \label{imin}
\end{equation}
In fact we have one more operator $T_L(1)$ which commutes with all $I_m$.

The usual choice for the Hamiltonian is
\begin{equation*}
H_L = I_1.
\end{equation*}
Assume that the initial condition
\begin{equation}
R_{V | V}(\zeta | \zeta) = c_V P_{V | V} \label{ic}
\end{equation}
is valid for some nonzero constant $c_V$. Here, as above, $P_{V | V}$ is the permutation operator on $V \otimes V$. One can demonstrate that in this case
\begin{equation}
H_L = \sum_{i \in \interval{1}{L}} \left. \frac{\rmd \check R_{V | V}(\zeta | 1)^{i, \, i + 1}}{\rmd \zeta} \right|_{\zeta = 1}, \label{hl}
\end{equation}
where we assume that
\begin{equation*}
\check R_{V | V}(\zeta | 1)^{L, \, L + 1} = \check R_{V | V}(\zeta | 1)^{L \, 1}.
\end{equation*}
Thus, we have a local Hamiltonian. For the well known simple graphical derivation of relation (\ref{hl}) we refer to paper \cite{NirRaz18}.

\subsection{Normalization}

In this paper we work with a fixed $\uqlg$-module $V$ and its dual $V^*$. We
choose the normalization of $R_{V^* | V} (\zeta | \eta)$, $R_{V | V^*} (\zeta
| \eta)$ and $R_{V^* | V^*}(\zeta | \eta)$ assuming that
\begin{gather*}
\rho_{V^* | V}(\zeta | \eta) = \rho_{V | V}(\zeta | \eta)^{-1}, \qquad \rho_{V | V^*}(\zeta | \eta) = \rho_{V | V}(\zeta | \eta)^{-1}, \\
\rho_{V^* | V^*}(\zeta | \eta) = \rho_{V | V}(\zeta | \eta).
\end{gather*}
In this case the crossing relation (\ref{ca}) implies
\begin{equation*}
R_{V^* | V}(\zeta | \eta) = (R_{V | V}(\zeta | \eta)^{-1})^{t_1}, \qquad R_{V^* | V^*}(\zeta | \eta) = (R_{V | V^*}(\zeta | \eta)^{-1})^{t_1}.
\end{equation*}
\begin{figure}[ht]
\centering
\begin{minipage}{0.4\textwidth}
\centering
\includegraphics{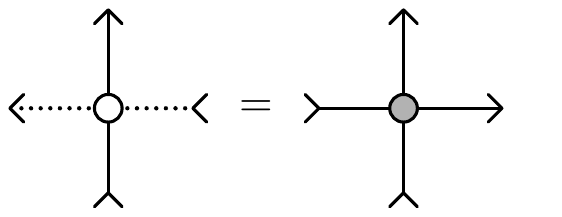}
\caption{}
\label{f:crani}
\end{minipage} \hfil
\begin{minipage}{0.4\textwidth}
\centering
\includegraphics{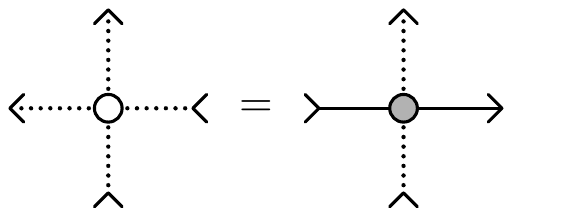}
\caption{}
\label{f:cranii}
\end{minipage}
\end{figure}
The graphical image of these relations can be found in figures \ref{f:crani} and \ref{f:cranii}. Here and below, for the representation $\varphi^*$ we use the dotted variant of the line used for the representation $\varphi$.

\begin{figure}[ht]
\includegraphics{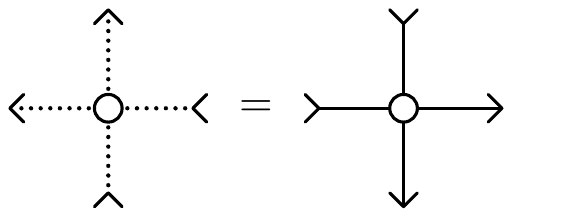}
\caption{}
\label{f:crbn}
\end{figure}

It is clear that the crossing relation (\ref{cb}) takes the form
\begin{equation}
R_{V^* | V^*}(\zeta | \eta) = R_{V | V}(\zeta | \eta)^t \label{crb}
\end{equation}
and has the graphical image given in figure \ref{f:crbn}.

\begin{figure}[ht]
\centering
\begin{minipage}{0.4\textwidth}
\centering
\includegraphics{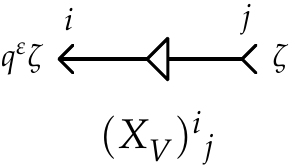}
\caption{}
\label{f:xo}
\end{minipage} \hfil
\begin{minipage}{0.4\textwidth}
\centering
\includegraphics{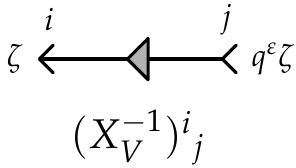}
\caption{}
\label{f:ixo}
\end{minipage}
\end{figure}

\begin{figure}[ht]
\centering
\begin{minipage}{0.4\textwidth}
\centering
\includegraphics{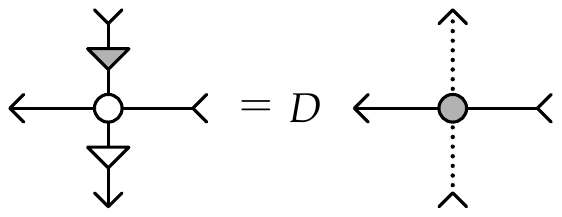}
\caption{}
\label{f:crcni}
\end{minipage} \hfil
\begin{minipage}{0.4\textwidth}
\centering
\includegraphics{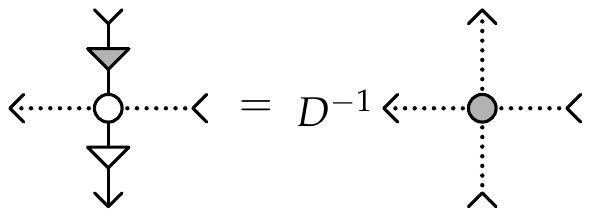}
\caption{}
\label{f:crcnii}
\end{minipage}
\end{figure}

Starting from the crossing relation (\ref{cc}), we obtain in the case under consideration two equations
\begin{gather}
(X_V \otimes 1_V) \, R_{V | V}(q^{-(\theta | \theta) \svee{h} / s} \zeta | \eta) \, (X_V^{-1} \otimes 1_V) = D(\zeta | \eta) \,  (R_{V^* | V}(\zeta | \eta)^{-1})^{t_1}, \label{ccni} \\
(X_V \otimes 1_{V^*}) \, R_{V | V^*}(q^{-(\theta | \theta) \svee{h} / s} \zeta | \eta) \, (X_V^{-1} \otimes 1_{V^*}) = D(\zeta | \eta)^{-1} \,  (R_{V^* | V^*}(\zeta | \eta)^{-1})^{t_1}, \label{ccnii}
\end{gather}
where
\begin{equation}
D(\zeta | \eta) = \rho_{V | V}(q^{- (\theta | \theta) \svee{h} / s} \zeta | \eta)^{-1} \, \rho_{V | V}(\zeta | \eta). \label{dze}
\end{equation}
To give a graphical interpretation of these equations, we use for the matrix elements of the operator $X_V$ and its inverse the depiction given in figures \ref{f:xo} and \ref{f:ixo}.
It can be demonstrated now that figures \ref{f:crcni} and \ref{f:crcnii} represent the crossing relations (\ref{ccni}) and (\ref{ccnii}).

\begin{figure}[ht]
\centering
\begin{minipage}[b]{0.4\textwidth}
\centering
\includegraphics{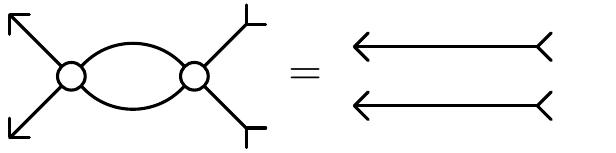}
\caption{}
\label{f:urani}
\end{minipage} \hfil
\begin{minipage}[b]{0.4\textwidth}
\centering
\includegraphics{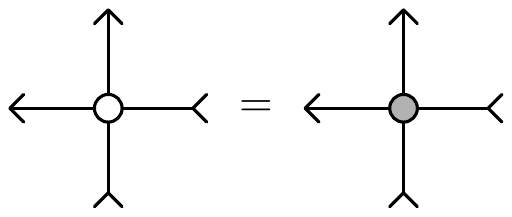}
\caption{}
\label{f:uranii}
\end{minipage}
\end{figure}

We choose the normalization factor so that the matrix elements of $R_{V | V}(\zeta | \eta)$ are rational functions of the spectral parameters, and $R_{V | V}(\zeta | \eta)$ satisfies the unitarity relation
\begin{equation}
\check R_{V | V}(\zeta | \eta) \check R_{V | V}(\eta | \zeta) = 1_{V \otimes V}, \label{urn}
\end{equation}
see \cite[Propositions 9.5.3 and 9.5.5]{EtiFreKir98}. We give the graphical form of this relation and the equivalent one in figures \ref{f:urani} and \ref{f:uranii}.

\begin{figure}[ht]
\centering
\begin{minipage}[b]{0.4\textwidth}
\centering
\includegraphics{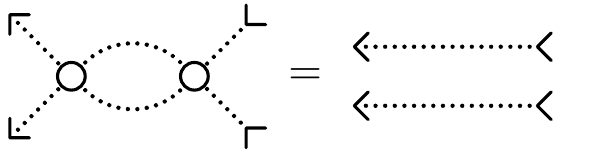}
\caption{}
\label{f:urbni}
\end{minipage} \hfil
\begin{minipage}[b]{0.4\textwidth}
\centering
\includegraphics{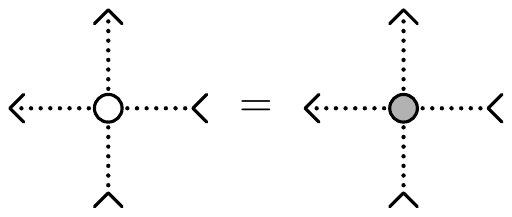}
\caption{}
\label{f:urbnii}
\end{minipage}
\end{figure}

The crossing relation (\ref{crb}) and the unitarity relation (\ref{urn}) lead to the unitarity relations in figures \ref{f:urbni} and \ref{f:urbnii}.

\begin{figure}[ht]
\centering
\includegraphics{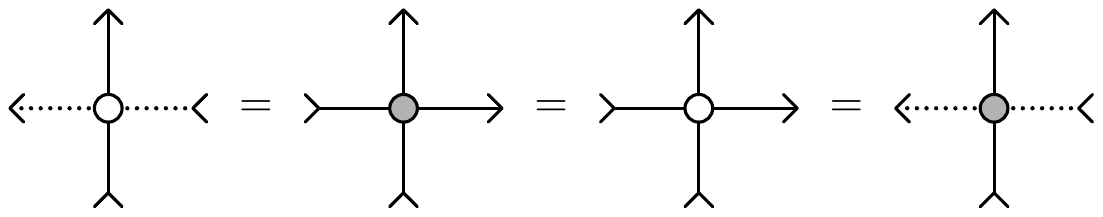}
\caption{}
\label{f:urcnp}
\end{figure}

\begin{figure}[ht]
\centering
\begin{minipage}{0.4\textwidth}
\centering
\includegraphics{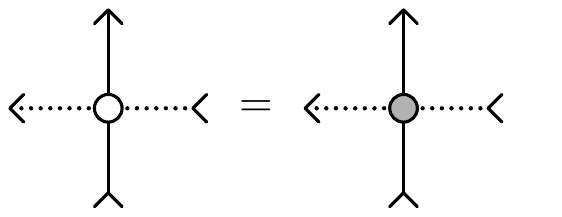}
\caption{}
\label{f:urcni}
\end{minipage} \hfil
\begin{minipage}{0.4\textwidth}
\centering
\includegraphics{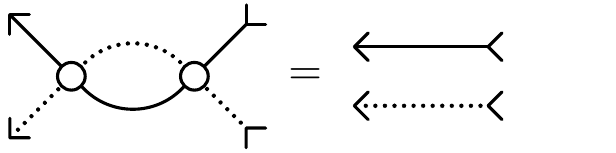}
\caption{}
\label{f:urcnii}
\end{minipage}
\end{figure}

Using the equations depicted in figures \ref{f:crani}, \ref{f:uranii} and \ref{f:crbn}, \ref{f:cranii} we come to the chain of equalities given in figure \ref{f:urcnp}.
We see that the $R$-operators $R_{V^* | V}(\zeta | \eta)$ and $R_{V | V^*}(\zeta | \eta)$ satisfy the unitarity relation given in figure \ref{f:urcni}, or the equivalent relation in figure \ref{f:urcnii}.
  
\begin{figure}[ht]
\centering
\begin{minipage}{0.4\textwidth}
\centering
\includegraphics{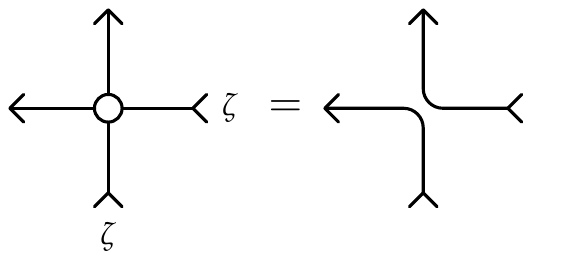}
\caption{}
\label{f:ici}
\end{minipage} \hfil
\begin{minipage}{0.4\textwidth}
\centering
\includegraphics{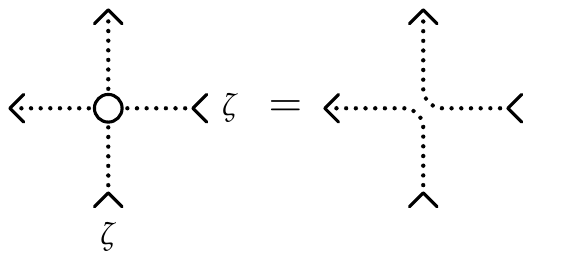}
\caption{}
\label{f:icii}
\end{minipage}
\end{figure}

Finally we assume that the initial condition (\ref{ic}) is satisfied. It follows from the unitarity relation (\ref{urn}) that in our case
\begin{equation*}
c_V^2 = 1.
\end{equation*}
Possibly changing the sign of $R_{V | V}(\zeta | \eta)$, without destroying the
form of the unitarity and crossing relations, we make $c_V$ equal to $1$. The
resulting initial condition is depicted in figure~\ref{f:ici}.
The crossing relation (\ref{cb}) has now the simple form
\begin{equation}
R_{V^* | V^*}(\zeta | \eta) = R_{V | V}(\zeta | \eta)^t, 
\end{equation}
and it leads to another initial condition represented by figure \ref{f:icii}.

\subsection{Density operator}

The density operator of a quantum statistical system with the Hamiltonian $H_L$ is given by the equation 
\begin{equation*}
D_L = \frac{1}{Z_L} \, \rme^{- \beta H_L}, \qquad \beta = \frac{1}{k T},
\end{equation*}
where $Z_L$ is the partition function of the system defined as
\begin{equation*}
Z_L = \tr \rme^{- \beta H_L}.
\end{equation*}
The expectation value of an arbitrary observable $F$ is
\begin{equation*}
\langle F \rangle = \tr (F D_L) = \frac{1}{Z_L} \tr (F  \, \rme^{- \beta H_L}).
\end{equation*}

Let us exploit the relation of the Hamiltonian $H_L$ with the transfer operator $T_L(\zeta)$. To this end we introduce the `additive' spectral parameter $u$ related to the `multiplicative' spectral parameter $\zeta$ by the relation
\begin{equation*}
q^u = \rme^{\hbar u} = \zeta.
\end{equation*}
Slightly abusing notation, we denote by $T_L(u)$ the transfer operator $T_L(\zeta)$ expressed as a function of $u$. Now we have
\begin{equation*}
I_m = \frac{1}{m! \, \hbar^m} \left. \frac{\rmd^m}{\rmd u^m} \, \log T_L(u) \right|_{u = 0},
\end{equation*}
and it is not difficult to see that
\begin{equation*}
T_L(u) = T_L(0) \exp \big(\sum_{m = 0}^\infty (\hbar \, u)^m I_m \big).
\end{equation*}

We consider one more transfer operator related to the module $V^*$ and defined as
\begin{equation*}
T^*_L(\zeta) = T_{V^* | V}(\zeta | \underbracket[.6pt]{1, \, 1, \, \ldots, \, 1}_L).
\end{equation*}
It generates one more set of commuting quantities
\begin{equation*}
I^*_m = \left. \Big(\zeta \frac{\rmd}{\rmd \zeta} \Big)^m \log T^*_L(\zeta) \right|_{\zeta = 1}. 
\end{equation*}
In fact, in addition to (\ref{imin}), we have
\begin{equation*}
[I^*_m, \, I^*_n] = 0, \qquad [I^{}_m, \, I^*_n] = 0, \qquad m, n \in \bbZ_{>0}.
\end{equation*}
The operators $T^{}_L(1)$ and $T^*_L(1)$ commute with all $I^{}_m$ and all $I^*_m$.
In fact, $T^{}_L(1)$ is the left shift and $T^*_L(1)$ is the right shift, and we have
\begin{equation*}
T^{}_L(1) T^*_L(1) = 1_{V^{\otimes L}}.
\end{equation*}
In terms of the additive spectral parameter $u$ we have
\begin{equation*}
T^*_L(u) = T^*_L(0) \, \exp \big( \sum_{m = 0}^\infty (\hbar \, u)^m I^*_m \big).
\end{equation*}

\begin{figure}[ht]
\includegraphics{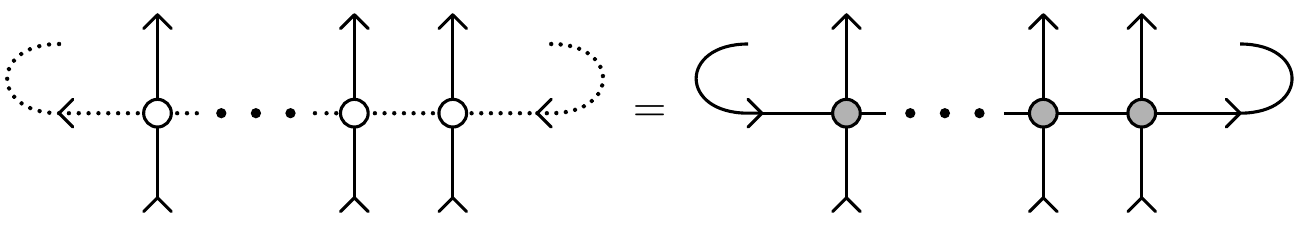}
\caption{}
\label{f:dto}
\end{figure}

Using the crossing relation given in figure \ref{f:crani}, we obtain the equation represented by figure \ref{f:dto}. 
Starting from this equation, we determine that
\begin{equation*}
I^*_1 = - I_1 = - H_L.
\end{equation*}

For any positive integer $N$ we can write
\begin{align*}
& T^{}_L(0)^{-N} \, T^{}_L(u/2 \hbar N)^N = \exp \Big(\frac{u}{2} I_1 + \sum_{m = 1}^\infty N^{- m} \big( \frac{u}{2} \big)^{m + 1} I_{m + 1} \Big), \\
& T^*_L(0)^{-N} \, T^*_L(-u/2\hbar N)^N = \exp \Big(- \frac{u}{2} I^*_1 - \sum_{m = 1}^\infty N^{- m} \big( \frac{u}{2} \big)^{m + 1} (-1)^m I^*_{m + 1} \Big).
\end{align*}
These equations give
\begin{multline*}
T^*_L(- u/2\hbar N)^N T^{}_L(u/2\hbar N)^N \\*
= \exp \Big( \frac{1}{2} u (I_1 - I_1^*) + \sum_{m = 1}^\infty N^{- m} \big( \frac{u}{2} \big)^{m + 1} (I_{m + 1} - (-1)^m I^*_{m + 1}) \Big),
\end{multline*}
and we see that
\begin{equation*}
\lim_{N \to \infty} \big( T^*_L (- u /2\hbar N) T^{}_L(u /2\hbar N) \big)^N = \exp(u H_L).
\end{equation*}
Denote
\begin{equation*}
D_{L, N} = \frac{(T^*_L (\beta /2 \hbar N) T^{}_L(- \beta /2 \hbar N))^N}{Z_{L, N}},
\end{equation*}
where
\begin{equation*}
Z_{L, N} = \tr (T^*_L (\beta / 2 \hbar N) T^{}_L(- \beta / 2 \hbar N))^N.
\end{equation*} 
Finally, using the multiplicative spectral parameter, we obtain
\begin{equation}
Z_{L, N} D_{L, N} = (T^*_L (q^{\beta/2 N}) T^{}_L(q^{- \beta/2 N}))^N. \label{zlndln}
\end{equation}
It is clear that
\begin{equation*}
D_L = \lim_{N \to \infty} D_{L, N}.
\end{equation*}

\subsection{Density operator as the partition function of a vertex model}

\begin{figure}[ht]
\centering
\includegraphics{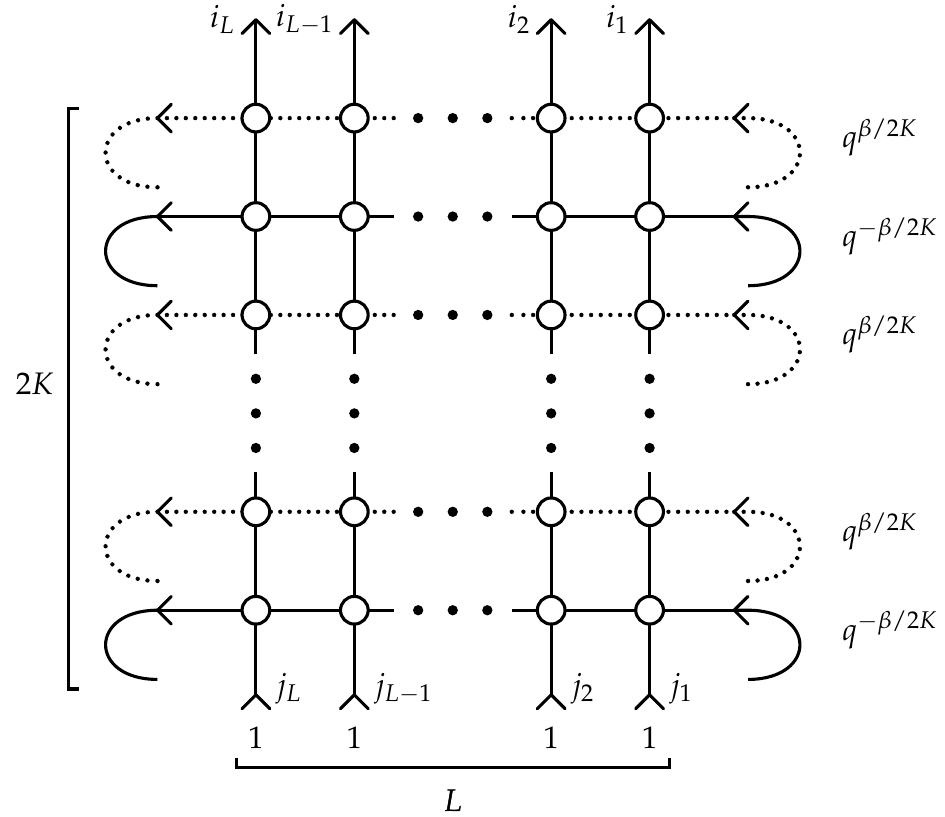}
\caption{}
\label{f:doln}
\end{figure}

It follows from (\ref{zlndln}) that the matrix elements of the operator $Z_{L, N} D_{L, N}$ can be represented as the partition function of a vertex model on a square lattice, see figure \ref{f:doln}.
Here we have the periodic boundary conditions in the horizontal direction and open top and bottom boundaries. The thermodynamic limit would be obtained when $L, N \to \infty$. However, the existence of the limit over $L$ is quite problematic. Therefore, we proceed to the density operator which allows to find expectation values only for local observables. To this end we assume that $L = 2 m + n$, where $m$ and $n$ are positive integers. We consider $n$ as fixed and take the trace of $Z_{L, N} D_{L, N}$ over the first and the last $m$ spaces associated with vertical directions of the lattice. We denote the corresponding `density operator' as $D_{n, N, m}$ and the corresponding `partition function' as $Z_{n, N, m}$. The density operator of interest is certainly the limit as $m \to \infty$ and $N \to \infty$. We assume that these limits commute, see a discussion in paper \cite{Suz85}, so that
\begin{equation*}
D_n = \lim_{m \to \infty} \lim_{N \to \infty} D_{n, N, m} = \lim_{N \to \infty} \lim_{m \to \infty} D_{n, N, m}.
\end{equation*}
To go further we generalize the objects under consideration in the following way.

\begin{figure}[ht]
\centering
\includegraphics{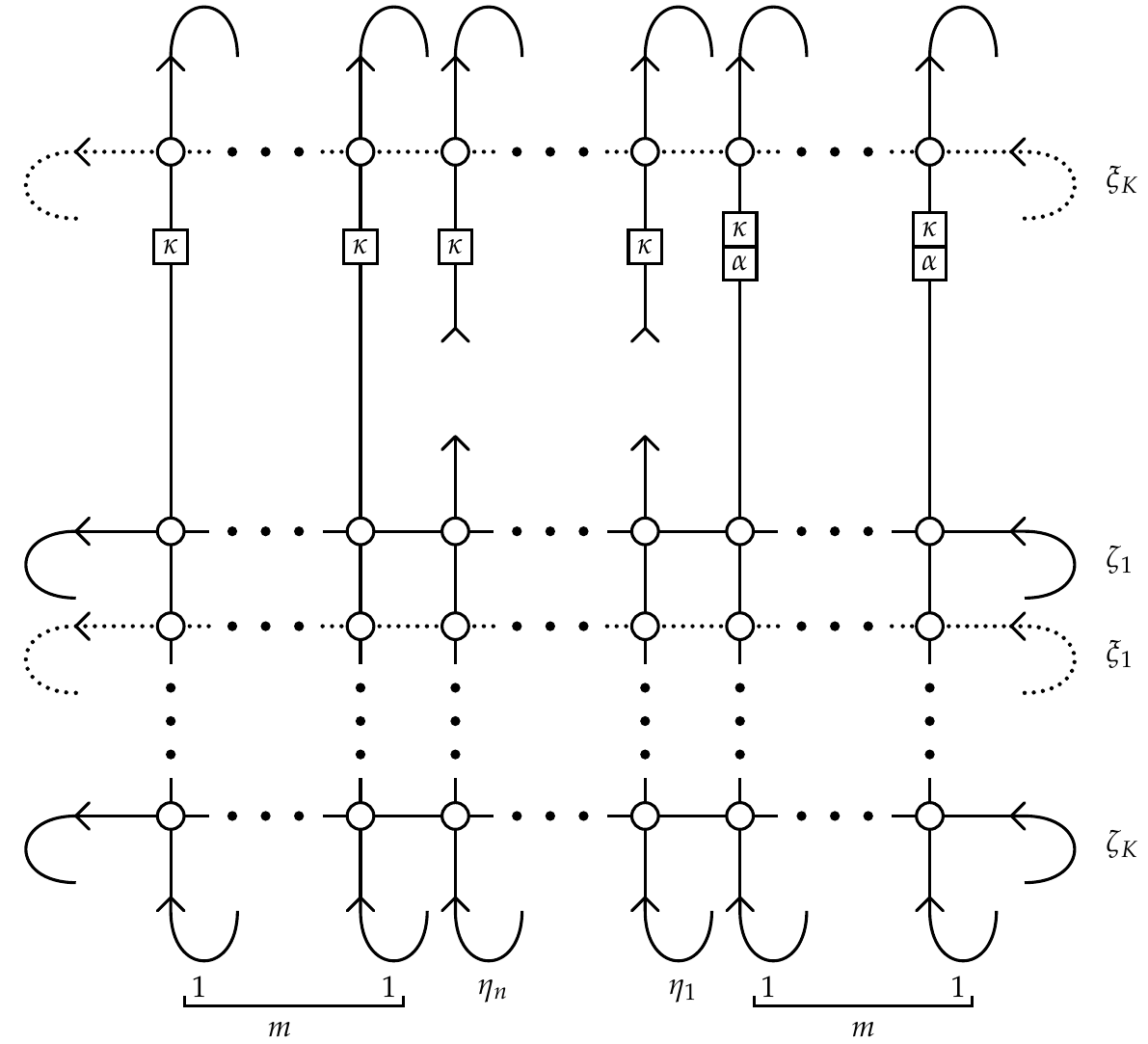}
\caption{}
\label{f:dolnn}
\end{figure}

We have horizontal transfer operators and vertical monodromy and transfer
operators defined in an evident way. We supply a horizontal transfer operator
with the spectral parameters $\zeta_1, \ldots, \zeta_N $ or $\xi_1, \ldots,
\xi_N$ in dependence on whether it is the operator $T$ or the operator $T^*$,
see figure \ref{f:dolnn}.
The vertical monodromy operators are endowed with the spectral parameters $\eta_1, \ldots, \eta_n$. Thus, we consider the generalized density operator
\begin{equation*}
D_{n, N, m}(\zeta_1, \ldots, \zeta_N, \xi_1, \ldots, \xi_N | \eta_1, \ldots, \eta_n).
\end{equation*}
Below, if it does not lead to misunderstanding, we omit the explicit designation of dependence on $\zeta_1, \ldots, \zeta_N$ and $\xi_1, \ldots, \xi_N$.

\begin{figure}[ht]
\centering
\begin{minipage}{0.4\textwidth}
\centering
\includegraphics{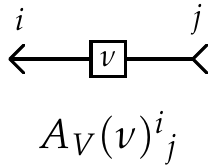}
\caption{}
\label{f:ao}
\end{minipage} \hfil
\begin{minipage}{0.4\textwidth}
\centering
\includegraphics{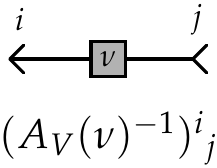}
\caption{}
\label{f:iao}
\end{minipage}
\end{figure}

\begin{figure}[ht]
\centering
\begin{minipage}{0.4\textwidth}
\centering
\includegraphics{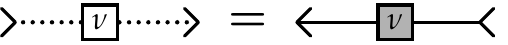}
\caption{}
\label{f:aca}
\end{minipage} \hfil
\begin{minipage}{0.4\textwidth}
\centering
\includegraphics{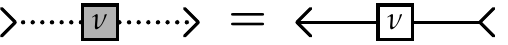}
\caption{}
\label{f:acb}
\end{minipage}
\end{figure}

After all we introduce some twisting for the vertical transfer and monodromy
operators. In the framework of the quantum group approach a twisting is
defined by a choice of a group-like element. Remember that an element $a$ of a
Hopf algebra is called group-like if
\begin{equation*}
\Delta(a) = a \otimes a.
\end{equation*}
It is clear that in our case an element
\begin{equation*}
a = q^{\sum_{i = 1}^l \nu_i h_i}
\end{equation*}
is group-like for any complex number $\nu_i$. We denote
\begin{equation*}
A(\nu) = \varphi(q^{\sum_{i = 1}^l \nu_i h_i})
\end{equation*}
and use for the matrix elements of the operator $A(\nu)$ and its inverse the depiction given in figures \ref{f:ao} and \ref{f:iao}.
One can demonstrate the validity of the graphical equations represented by figures \ref{f:aca} and \ref{f:acb}.

\begin{figure}[ht]
\centering
\begin{minipage}{0.45\textwidth}
\centering
\includegraphics{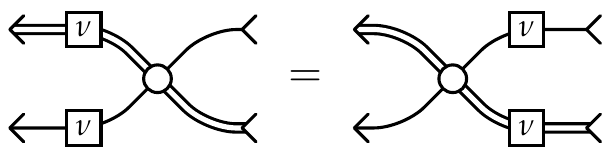}
\caption{}
\label{f:roao}
\end{minipage} \hfil
\begin{minipage}{0.45\textwidth}
\centering
\includegraphics{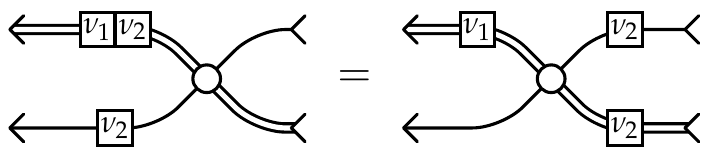}
\caption{}
\label{f:roaom}
\end{minipage}
\end{figure}

\begin{figure}[ht]
\centering
\includegraphics{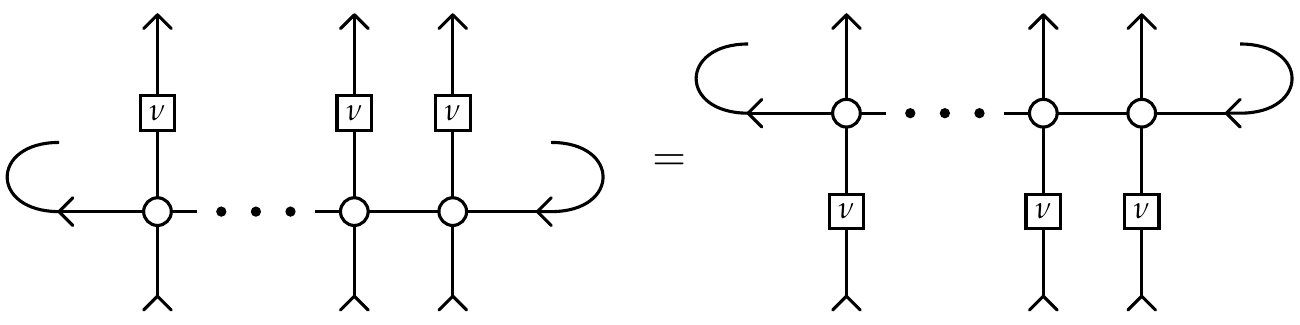}
\caption{}
\label{f:ctoa}
\end{figure}

It follows from the definition of a group-like element that the operator
$A_V(\nu)$ satisfies a useful equation whose graphical image is represented by
figure \ref{f:roao}.
It is also clear that
\begin{equation*}
A_V(\nu_1 + \nu_2) = A_V(\nu_1) A_V(\nu_2) = A_V(\nu_2) A_V(\nu_1).
\end{equation*}
This relation leads to a modified version of the graphical equation \ref{f:roao} which can be seen in figure \ref{f:roaom}. We also need the commutativity equation given in figure \ref{f:ctoa}.

We introduce disorder parameters $\alpha_1$, $\ldots$, $\alpha_l$ and twist the
first $m$ vertical transfer operators. The introduction of disorder parameters
regularizes the problem in the case of $\uqlslii$ \cite{BooJimMiwSmiTak07, BooJimMiwSmiTak09}. Further, we introduce parameters $\kappa_1$, $\ldots$,
$\kappa_l$ and twist all vertical transfer and monodromy operators, see figure
\ref{f:dolnn}. This can be interpreted as turning on a `magnetic field'. It
should be noted that all equally twisted transfer operators commute.

Denote by $\calV$ the horizontal space,
\begin{equation*}
\calV = \underbracket[.6pt]{V \otimes V^* \otimes \cdots \otimes V \otimes V^*}_{2 N}. 
\end{equation*}
We use for a vertical monodromy operator, twisted with the parameters $\nu_1$, $\ldots$, $\nu_l$ the notation $\calM^\nu(\zeta_1, \xi_1, \ldots, \zeta_N, \xi_N | \eta)$. It acts on the space $\calV \otimes V$. In fact, we have
\begin{multline*}
\calM^\nu(\zeta_1, \xi_1, \ldots, \zeta_N, \xi_N | \eta) \\*
= \big( ((\varphi^{}_{\zeta_1} \otimes_{\Delta'} \varphi^*_{\xi_1} \otimes_{\Delta'} \cdots \otimes_{\Delta'} \varphi^{}_{\zeta_N} \otimes_{\Delta'} \varphi^*_{\xi_N}) \otimes \varphi_\eta \big)(\calR)) (1_\calV \otimes A_V(\nu)).
\end{multline*}
It is useful to represent a vertical monodromy operator as
\begin{equation*}
\calM(\zeta_1, \xi_1, \ldots, \zeta_N, \xi_N | \eta) = \sum_{i, j} \calM(\zeta_1, \xi_1, \ldots, \zeta_N, \xi_N | \eta)^i{}_j \, E^j{}_i,
\end{equation*}
where $E^j{}_i$ are unit operators on $V$ associated with the used basis, and the operators $ \calM(\eta)^i{}_j$ act on $\calV$. The vertical transfer operator $\calT(\eta)$ is defined as
\begin{multline*}
\calT(\zeta_1, \xi_1, \ldots, \zeta_N, \xi_N | \eta) = \tr_V \calM(\zeta_1, \xi_1, \ldots, \zeta_N, \xi_N | \eta) = \sum_i \calM(\zeta_1, \xi_1, \ldots, \zeta_N, \xi_N | \eta)^i{}_i.
\end{multline*}
It acts on the vertical space $\calV$. We extend to the vertical monodromy and transfer operators the convention to omit the explicit designation of dependence on $\zeta_1, \ldots, \zeta_N$ and $\xi_1, \ldots, \xi_N$.

Looking at figure \ref{f:dolnn}, it is easy to see that
\begin{equation*}
D_{n, N, m}(\eta_1, \ldots, \eta_n)^{i_1 \ldots i_n}{}_{j_1 \ldots j_n}  = \frac{\tr \big( (\calT^\kappa)^m \calM^{\kappa}(\eta_n)^{i_n}{}_{j_n} \ldots \calM^{\kappa}(\eta_1)^{i_1}{}_{j_1} (\calT^{\kappa + \alpha})^m \big)}{\tr \big( (\calT^\kappa)^m \calT^{\kappa}(\eta_n) \ldots \calT^{\kappa}(\eta_1) (\calT^{\kappa + \alpha})^m \big)}.
\end{equation*}
Here and below we write instead of $\calT^\nu(1)$ just $\calT^\nu$.

Generalizing the conjecture made in \cite{JimMiwSmi09}, we assume that the transfer operators $\calT^\kappa(\eta)$ and $\calT^{\kappa + \alpha}(\eta)$ are diagonalizable. Due to the commutativity of the vertical transfer operators $\calT^\nu(\eta)$ with different spectral parameters $\eta$, their
eigenvectors can be chosen independently of $\eta$. Let the eigenvectors $v^\nu_a$ of $\calT^\nu(\eta)$ form a basis of the vertical space. We have
\begin{equation*}
\calT^\nu(\eta)v^\nu_a  = \lambda^\nu_a(\eta) \, v^\nu_a,
\end{equation*}
where $\lambda^\nu_a(\eta)$ are the corresponding eigenvalues. Denote by $\psi^\nu_a$ the vectors forming the dual basis, so that
\begin{equation*}
\langle \psi^\nu_a, \, v^\nu_b \rangle = \delta_{a b}.
\end{equation*}
Now we have
\begin{align*}
& \tr \big( (\calT^\kappa)^m \calM^{\kappa}(\eta_n)^{i_n}{}_{j_n} \ldots \calM^{\kappa}(\eta_1)^{i_1}{}_{j_1} (\calT^{\kappa + \alpha})^m \big) \\
& \hspace{4em} {} = \sum_a \langle \psi^{\kappa + \alpha}_a, \, (\calT^\kappa)^m \calM^{\kappa}(\eta_n)^{i_n}{}_{j_n} \ldots \calM^{\kappa}(\eta_1)^{i_1}{}_{j_1}(\calT^{\kappa + \alpha})^m \, v^{\kappa + \alpha}_a \rangle \\
& \hspace{4em} {} = \sum_{a, b} \langle \psi^{\kappa + \alpha}_a, \, (\calT^\kappa)^m v^\kappa_b \rangle \langle \psi^\kappa_b, \,  \calM^{\kappa}(\eta_n)^{i_n}{}_{j_n} \ldots \calM^{\kappa}(\eta_1)^{i_1}{}_{j_1}(\calT^{\kappa + \alpha})^m \, v^{\kappa + \alpha}_a \rangle \\
& \hspace{8em} {} = \sum_{a, b} (\lambda^{\kappa + \alpha}_a)^m (\lambda^\kappa_b)^m  \langle \psi^{\kappa + \alpha}_a, \, v^\kappa_b \rangle \langle \psi^\kappa_b, \,  \calM^{\kappa}(\eta_n)^{i_n}{}_{j_n} \ldots \calM^{\kappa}(\eta_1)^{i_1}{}_{j_1} \, v^{\kappa + \alpha}_a \rangle,
\end{align*}
where instead of $\lambda^\nu_a(1)$ we write just  $\lambda^\nu_a$. In a similar way we obtain
\begin{multline*}
\tr \big( (\calT^\kappa)^m \calT^{\kappa}(\eta_n) \ldots \calT^{\kappa}(\eta_1) (\calT^{\kappa + \alpha})^m \big) \\
{} = \sum_{a, b} (\lambda^{\kappa + \alpha}_a)^m (\lambda^\kappa_b)^m  \lambda^\kappa_b(\eta_n) \ldots \lambda^\kappa_b(\eta_1) \langle \psi^{\kappa + \alpha}_a, \, v^\kappa_b \rangle \langle \psi^\kappa_b, \, v^{\kappa + \alpha}_a \rangle.
\end{multline*}
Following again paper \cite{JimMiwSmi09}, we assume that the eigenvalues $\lambda^\kappa_0$ and $\lambda^{\kappa + \alpha}_0$ of $\calT^\kappa$ and $\calT^{\kappa + \alpha}$ with the maximal absolute value are non-degenerate. In this case in the limit $m \to \infty$ we get
\begin{equation*}
D_{n, N}(\eta_1, \ldots, \eta_n)^{i_1 \ldots i_n}{}_{j_1 \ldots j_n}  = \frac{\langle \psi^\kappa_0, \,  \calM^{\kappa}(\eta_n)^{i_n}{}_{j_n} \ldots \calM^{\kappa}(\eta_1)^{i_1}{}_{j_1} \, v^{\kappa + \alpha}_0 \rangle}{\lambda^\kappa_0(\eta_n) \ldots \lambda^\kappa_0(\eta_1) \langle \psi^\kappa_0, \, v^{\kappa + \alpha}_0 \rangle}.
\end{equation*}
Here we assume also that
\begin{equation*}
\langle \psi^\kappa_0, \, v^{\kappa + \alpha}_0 \rangle \ne 0, \qquad \langle \psi^{\kappa + \alpha}_0, \, v^\kappa_0 \rangle \ne 0.
\end{equation*}

\section{Reduced qKZ equation} \label{s:rqe}

In this section we describe a graphical derivation of the discrete reduced qKZ
equation for an arbitrary quantum loop algebra and consider the zero
temperature limit. For the case of $\uqlslii$ this was done in the
thesis~\cite{Auf11}, see also \cite{AufKlu12}. The case of $\uqlsliii$ was
treated in \cite{BooHutNir18} and, using an alternative approach, in
\cite{RibKlu19}. It appears that in the general case it is convenient to split the
equation into two equations and consider them separately.

\subsection{First equation}

The graphical derivation of the first equation is given in figures
\ref{f:drqkza}-\ref{f:drqkzi} with appropriate comments. The sought
equation arises from comparison of figures \ref{f:drqkza} and
\ref{f:drqkzi}. Looking at figure \ref{f:drqkzi}, we see that it is
constructive to generalize the concept of density operator. Namely, new
operators are also described by the picture similar to \ref{f:dolnn}. However, some
vertical cut lines can be associated with the dual representation $\varphi^*$,
which is reflected by using a dotted line. We denote the corresponding
monodromy and transfer operators as $\calM^{* \nu}(\eta)$ and $\calT^{*
  \nu}(\eta)$. To be more precise, we illustrate our definition by the
following analytical expression
\begin{multline*}
D_{n, N}(\eta_1, \ldots, \eta^*_k, \ldots, \eta_n)^{i_1 \ldots i_k \ldots i_n}{}_{j_1 \ldots j_k \ldots j_n}  \\*
= \frac{\langle \psi^\kappa_0, \,  \calM^{\kappa}(\eta_n)^{i_n}{}_{j_n} \ldots \calM^{*\kappa}(\eta_k)^{i_k}{}_{j_k} \ldots \calM^{\kappa}(\eta_1)^{i_1}{}_{j_1} \, v^{\kappa + \alpha}_0 \rangle}{\lambda^\kappa_0(\eta_n) \ldots \lambda^{* \kappa}_0(\eta_k) \ldots \lambda^\kappa_0(\eta_1) \langle \psi^\kappa_0, \, v^{\kappa + \alpha}_0 \rangle}.
\end{multline*}
Here we use for the corresponding spectral parameter $\eta_i$ the notation $\eta^*_i$, having in mind that it is actually $\eta_i$ but associated with the dual representation. Using the commutativity of the vertical transfer matrices $\calT^\nu(\eta)$ and $\calT^{* \nu}(\eta)$, we assume that $v^\nu_a$ are also eigenvectors of $\calT^{* \nu}(\eta)$ and mark the corresponding eigenvalues by an asterisk, so that
\begin{equation*}
\calT^{* \nu}(\eta) v^\nu_a = \lambda^{* \nu}_a(\eta) v^\nu_a.
\end{equation*}

If we take the operator graphically described by figure \ref{f:drqkza}, divide it by the scalar $Z_{n, N, m}(\eta_1, \ldots, \eta_n)$, put $\eta_n = \zeta_1$ and take the limit $m \to \infty$, we obtain the action of some linear operator $A_n(\eta_1, \ldots, \eta_{n - 1}, \zeta_1)$ on the operator $D_{n, N}(\eta_1, \ldots, \eta_{n - 1}, \zeta_1)$. Applying this procedure to the operator given in figure~\ref{f:drqkzi}, we come to the expression
\begin{multline*}
\lambda^\kappa_0(\zeta_1) \lambda^{\kappa + \alpha}_0(\zeta_1) \frac{\langle \psi^\kappa_0, \,  \calM^{* \kappa}(q^\lambda \zeta_1)^{i_n}{}_{j_n} \calM^\kappa(\eta_{n - 1})^{i_{n - 1}}{}_{j_{n - 1}} \ldots \calM^{\kappa}(\eta_1)^{i_1}{}_{j_1} \, v^{\kappa + \alpha}_0 \rangle}{\lambda^\kappa_0(\zeta_1) \lambda^\kappa_0(\eta_{n - 1}) \ldots \lambda^\kappa_0(\eta_1)  \langle \psi^\kappa_0, \, v^{\kappa + \alpha}_0 \rangle} \\
= \lambda^{\kappa + \alpha}_0(\zeta_1) \lambda^{* \kappa}_0 (q^\lambda \zeta_1) D_{n, N}(\eta_1, \ldots, \eta_{n -1}, (q^\lambda \zeta_1)^*).
\end{multline*}
It is worth to remind here that $\lambda = (\theta | \theta) \svee{h}/s$.

\begin{figure}[ht]
\centering
\includegraphics{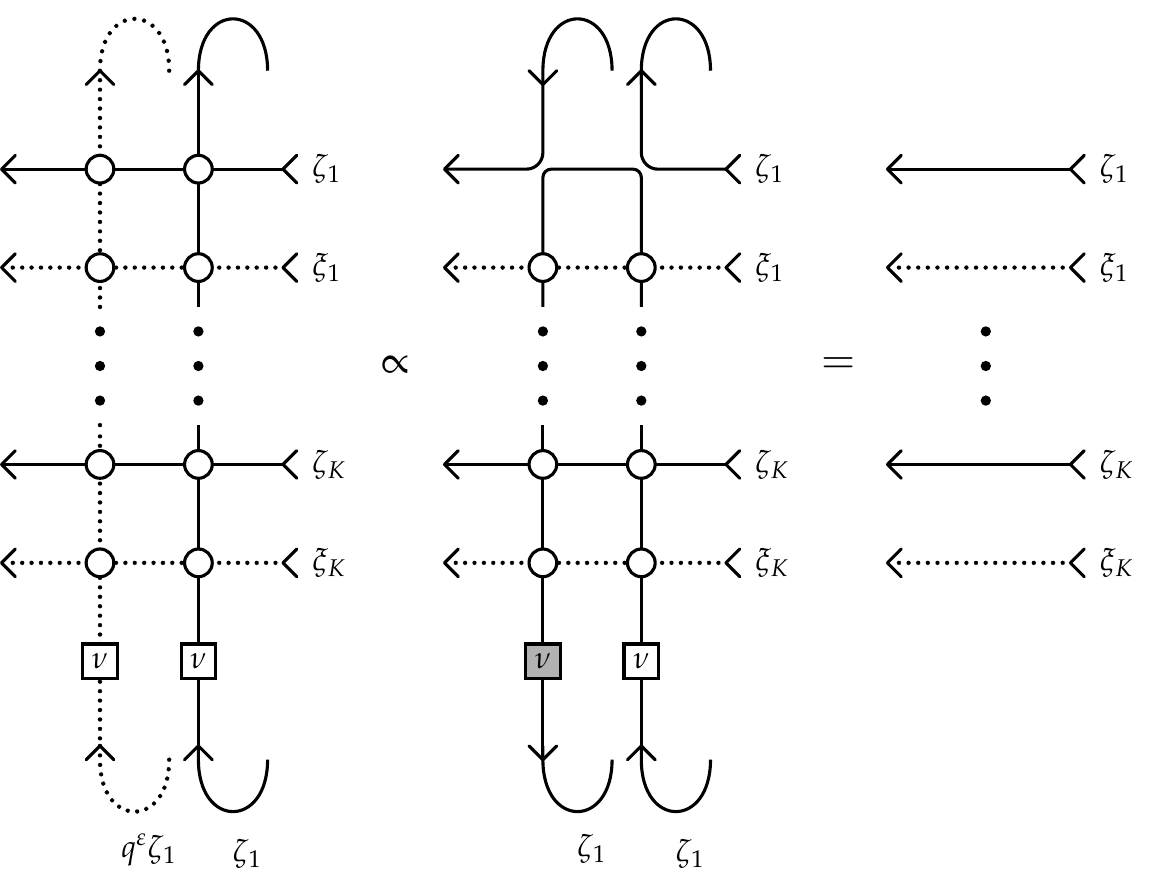}
\caption{}
\label{f:tosto}
\end{figure}

Consider now the product $\calT^{* \nu}(q^\lambda \zeta_1) \calT^\nu(\zeta_1)$. It is represented by the left picture in figure \ref{f:tosto}.
Successively applying the crossing relations \ref{f:crcni}, \ref{f:crcnii} and \ref{f:aca}, the unitarity relations~\ref{f:uranii} and~\ref{f:urcni}, the commutativity of the operators $X_V$ and $A_V(\nu)$, and the initial condition \ref{f:ici}, we come to the middle picture. Here we acquire the scalar factor $\prod_{i = 1}^N D^{-1}(q^\lambda \zeta_1 | \zeta_i) D(q^\lambda \zeta_1 | \xi_i)$. Finally, using the unitarity relations~\ref{f:urani} and~\ref{f:urcnii}, we get the right picture. Thus, we have the equation
\begin{equation}
\calT^{* \nu}(q^\lambda \zeta_1) \calT^\nu(\zeta_1) = \Big( \prod_{i = 1}^N D^{-1}(q^\lambda \zeta_1 | \zeta_i) D(q^\lambda \zeta_1 | \xi_i) \Big) \, 1_\calV, \label{tst}
\end{equation}
or in terms of eigenvalues
\begin{equation*}
\lambda^{* \nu}_a(q^\lambda \zeta_1) \lambda^\nu_a(\zeta_1) = \prod_{i = 1}^N D^{-1}(q^\lambda \zeta_1 | \zeta_i) D(q^\lambda \zeta_1 | \xi_i).
\end{equation*}
Remembering now about the factor we acquired in transition from figure \ref{f:drqkza} to figure \ref{f:drqkzi}, we conclude that the comparison of these figures gives the equation
\begin{multline}
\frac{\lambda^\kappa_0 (\zeta_1)}{\lambda^{\kappa + \alpha}_0(\zeta_1)} A_n(\eta_1, \ldots, \eta_{n - 1}, \zeta_1) (D_{n, N} (\eta_1, \ldots, \eta_{n - 1}, \zeta_1)) \\=  D_{n, N}(\eta_1, \ldots, \eta_{n -1}, (q^\lambda \zeta_1)^*). \label{andnf}
\end{multline}
This equation, together with (\ref{andnf}), can be used, in particular, for investigation of the correlation functions at finite non-zero temperature. However, the necessity to fix some spectral parameters leads to some problems and the additional work is required. Here we consider the zero temperature limit which is obtained as follows. We put $\zeta_i = q^{-\beta/2N}$ and $\xi_i = q^{\beta/2N}$ and take the limit $\beta \to \infty$, $N \to \infty$, keeping the ratio $\beta/N$ fixed and equal to $-2\log \eta_n/\hbar$. The resulting equation is
\begin{equation}
\phi(\eta_n) \,  A_n(\eta_1, \ldots, \eta_{n - 1}, \eta_n) (D_{n, N} (\eta_1, \ldots, \eta_{n - 1}, \eta_n)) = D_{n, N}(\eta_1, \ldots, \eta_{n -1}, (q^\lambda \eta_n)^*), \label{andn}
\end{equation}
where
\begin{equation*}
\phi(\eta) = \lim_{N \to \infty}  \lambda^\kappa_0(\eta |  \underbracket[.6pt]{\eta, \, \eta^{-1}, \, \ldots, \, \eta, \, \eta^{-1}}_{2N}) / \lambda^{\kappa + \alpha}_0(\eta |  \underbracket[.6pt]{\eta, \, \eta^{-1}, \, \ldots, \, \eta, \, \eta^{-1}}_{2N}).
\end{equation*}
Certainly, in the case $\alpha_i = 0$, $i \in \interval{1}{l}$, we have $\phi(\eta) = 1$.

\begin{figure}[ht]
\centering
\includegraphics{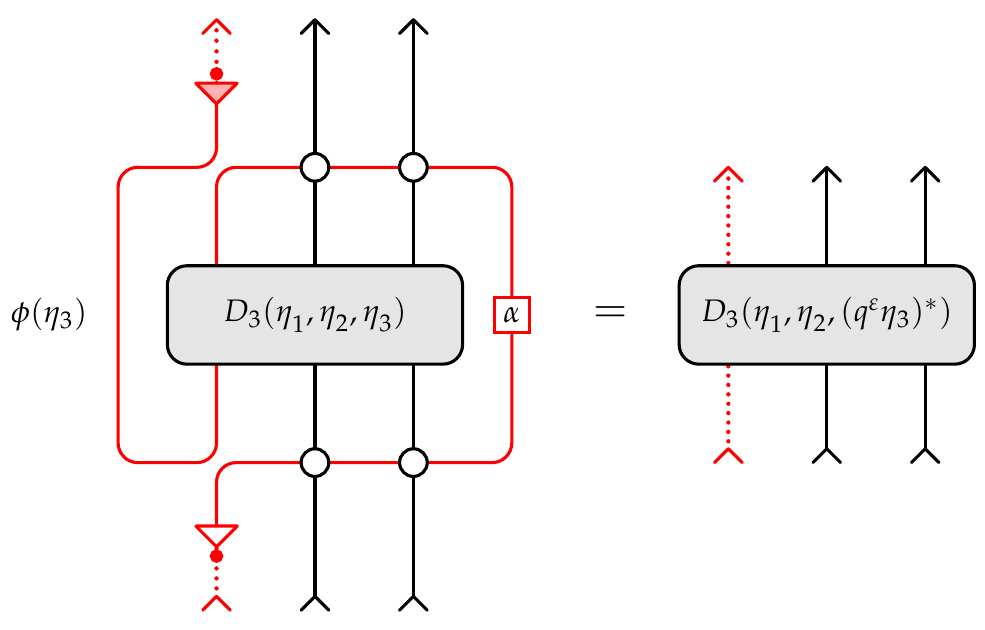}
\caption{}
\label{f:fe}
\end{figure}

\begin{figure}[ht]
\centering
\begin{minipage}[b]{0.4\textwidth}
\centering
\includegraphics{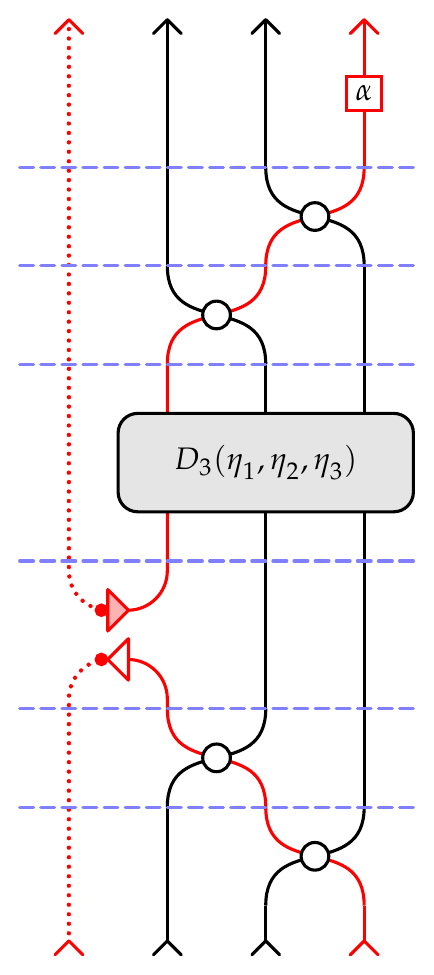}
\caption{}
\label{f:fedfrmd}
\end{minipage} \hfil
\begin{minipage}[b]{0.4\textwidth}
\centering
\includegraphics{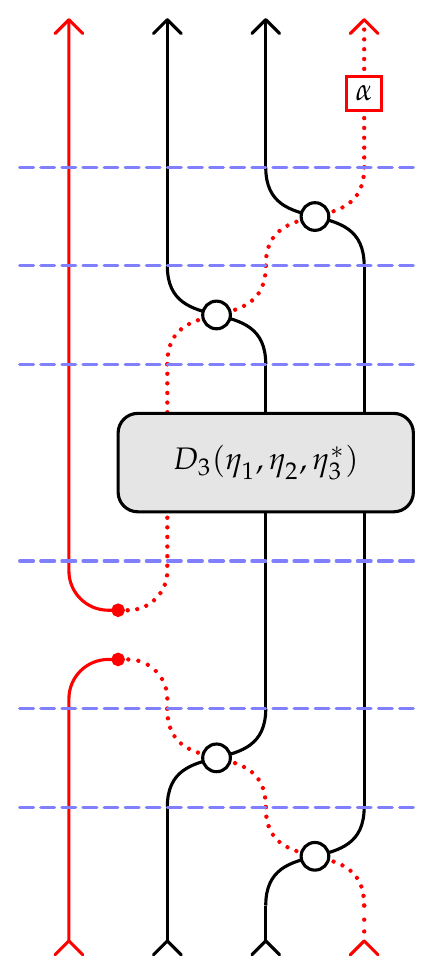}
\caption{}
\label{f:sedfrmd}
\end{minipage}
\end{figure}

It is constructive to give the graphical image of equation (\ref{andn}). Below, using pictures, we assume that $n = 3$. It is enough to understand the general situation. It is clear that figure \ref{f:fe} depicts equation (\ref{andn}).
Fat dots in the picture means changing the interpretation of the type of line. Namely, an input line corresponding to a representation is treated as the output line corresponding to the dual representation and so on. Note that the order of the vector spaces is from the right to the left. Cut the red line above the box with the label $\alpha$ in the left hand side of this equation and slightly deform the picture to obtain figure \ref{f:fedfrmd}.

Remember that if $V$ is finite dimensional, the space $\End(V)$ of linear
operators on $V$ can be identified with the space $V \otimes V^*$. To this end
one defines the mapping $\iota_V \colon V \otimes V^* \to \End(V)$ by the
equation
\begin{equation*}
\iota_V(v \otimes \psi) \, u = v \, \langle \psi, \, u \rangle.
\end{equation*}
One can show that it is a bijective mapping. Using it, one defines the mapping
from $\End(V)$ to $\End(V \otimes V^*)$ which sends an operator $F$ to the
operator
\begin{equation*}
\bbF = \iota^{}_{V \otimes V^*} \big( \iota^{-1}_V(F^{-1}) \otimes \iota^{-1}_{V^*}(F^t) \big).
\end{equation*}
We numerate the vector spaces as $0, 1, \ldots, n$. Now we can write the analytical expression for the figure \ref{f:fedfrmd}. It is not difficult to generalize it to the case of an arbitrary $n$. Now, taking the trace over the additional space, we come to the following analytical expression for the first equation
\begin{multline}
\phi(\eta_n) \,  \tr_0 \big( A_V^{(0)}(\alpha) \check R^{(01)}_{V | V}(\eta_1 | \eta_n) \ldots \check R^{(n - 2, n - 1)}_{V | V}(\eta_{n - 1} | \eta_n) \\*
D^{(0, 1, \ldots,  n - 1)}_n(\eta^{}_1, \ldots, \eta^{}_{n - 1}, \eta^{}_n) \\*
\mathbbm{X}^{(n - 1, n)}_V \check R^{(n -2, n - 1)}_{V | V}(\eta_n | \eta_{n - 1}) \ldots \check R^{(01)}_{V | V}(\eta_n | \eta_1) \big) = D_n(\eta^{}_1, \ldots \eta^{}_{n - 1}, (q^\lambda \eta_n)^*). \label{fea}
\end{multline}

\subsection{Second equation}

The graphical proof of the second equation is very similar to the proof of the first one. The initial and the final points can be found in figures \ref{f:drqkzj} and \ref{f:drqkzr}. If we take the operator depicted in figure \ref{f:drqkzj}, divide it by $Z_{n, N, m}(\eta_1, \ldots, \eta_{n - 1}, \eta^*_n)$, put $\eta_n = \xi_1$ and take the limit $m \to \infty$, we obtain the action of some linear operator $B_n(\eta_1, \ldots, \eta_{n - 1}, \xi^*_1)$ on the operator $D_{n, N}(\eta_1, \ldots, \eta_{n - 1}, \xi^*_1)$. Applying this procedure to the operator of figure \ref{f:drqkzr}, we come to the expression
\begin{multline*}
\lambda^{*\kappa}_0(\xi_1) \lambda^{*, \, \kappa + \alpha}_0(\xi_1) \frac{\langle \psi^\kappa_0, \,  \calM^\kappa(\xi_1)^{i_n}{}_{j_n} \calM^\kappa(\eta_{n - 1})^{i_{n - 1}}{}_{j_{n - 1}} \ldots \calM^{\kappa}(\eta_1)^{i_1}{}_{j_1} \, v^{\kappa + \alpha}_0 \rangle}{\lambda^{*\kappa}_0(\xi_1) \lambda^\kappa_0(\eta_{n - 1}) \ldots \lambda^\kappa_0(\eta_1)  \langle \psi^\kappa_0, \, v^{\kappa + \alpha}_0 \rangle} \\
= \lambda^{*, \, \kappa + \alpha}_0(\zeta_1) \lambda^\kappa_0 (\xi_1) D_{n, N}(\eta_1, \ldots, \eta_{n -1}, \xi_1).
\end{multline*}
In a similar way as for equation (\ref{tst}) we obtain
\begin{equation*}
\calT^\nu(\xi_1) \calT^{* \nu}(\xi_1) = 1_\calV,
\end{equation*}
or in terms of eigenvalues
\begin{equation*}
\lambda^\nu_a(\xi_1) \lambda^{* \nu}_a(\xi_1) = 1.
\end{equation*}
Using this relation, we see that the comparison of figures \ref{f:drqkzj} and \ref{f:drqkzr} leads to the equation
\begin{equation}
\frac{\lambda^{* \kappa}_0 (\zeta_1)}{\lambda^{*, \, \kappa + \alpha}_0(\zeta_1)} B_n(\eta_1, \ldots, \eta_{n - 1}, \xi_1) D_{n, N} (\eta_1, \ldots, \eta_{n - 1}, \xi^*_1) =  D_{n, N}(\eta_1, \ldots, \eta_{n -1}, \xi_1). \label{bndnf}
\end{equation}
The zero temperature limit is obtained as follows. We put $\zeta_i = q^{-\beta/2N}$ and $\xi_i = q^{\beta/2N}$ and take the limit $\beta \to \infty$, $N \to \infty$, keeping the ratio $\beta/N$ fixed and equal to $2\log \eta_n/\hbar$. The resulting equation is
\begin{equation}
\phi^*(\eta_n) \, B_n(\eta_1, \ldots,\eta_n) (D_{n, N} (\eta_1, \ldots, \eta_n)) = D_{n, N}(\eta_1, \ldots, \eta_n), \label{bndn}
\end{equation}
where
\begin{equation*}
\phi^*(\eta) = \lim_{N \to \infty} \lambda^{* \kappa}_0(\eta |  \underbracket[.6pt]{\eta^{-1}, \, \eta, \, \ldots, \, \eta^{-1}, \, \eta}_{2N}) / \lambda^{*, \, \kappa + \alpha}_0(\eta |  \underbracket[.6pt]{\eta^{-1}, \, \eta, \, \ldots, \, \eta^{-1}, \, \eta}_{2N}).
\end{equation*}
In the case $\alpha_i = 0$, $i \in \interval{1}{l}$, we have $\phi^*(\eta) = 1$.

\begin{figure}[ht]
\centering
\includegraphics{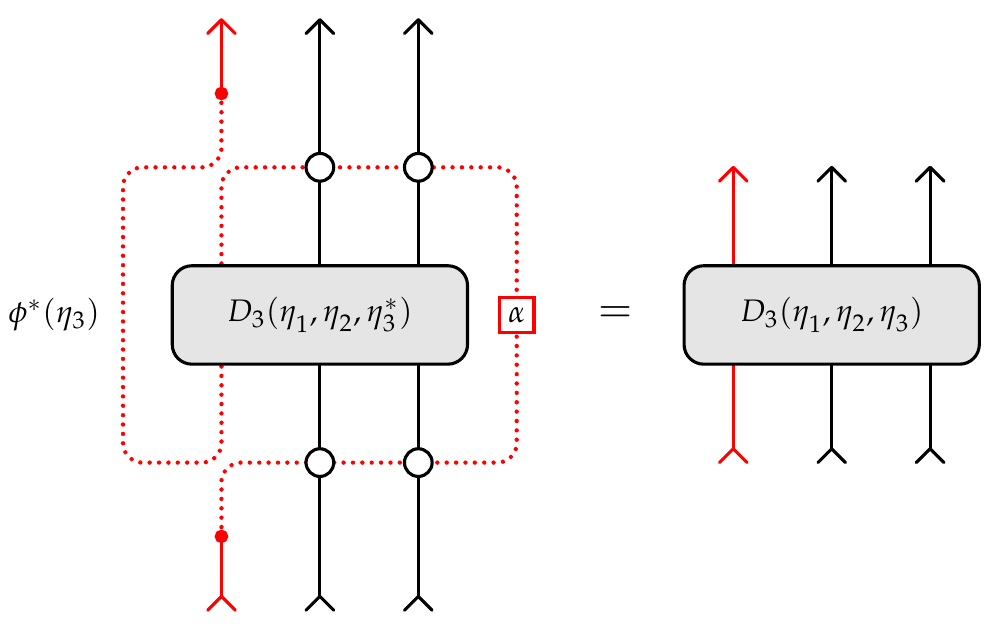}
\caption{}
\label{f:se}
\end{figure}

It is clear that figure \ref{f:se} depicts equation (\ref{bndn}).
Cut the dotted red line in the left hand side of this equation above the box with the label $\alpha$ and slightly deform the picture to obtain figure \ref{f:sedfrmd}. Writing the analytical expression for the figure \ref{f:sedfrmd}, generalizing to the case of an arbitrary $n$, and taking the trace over the additional space, come to the following analytical expression for the second equation
\begin{multline}
\phi^*(\eta_n) \, \tr_0 \big( A^{(0)}_{V^*}(\alpha) \check R^{(01)}_{V | V^*}(\eta_1 | \eta_n) \ldots \check R^{(n - 2, n - 1)}_{V | V^*}(\eta_{n - 1} | \eta_n) \\*
D^{(0, 1, \ldots, n - 1)}_n(\eta^{}_1, \ldots \eta^{}_{n - 1}, \eta^*_n) \\*
\mathbbm{1}_{V^*} \check R^{(n - 2, n - 1)}_{V^* | V}(\eta_n | \eta_{n - 1}) \ldots \check R^{(01)}_{V^* | V}(\eta_n | \eta_1) \big) = D_n(\eta^{}_1, \ldots, \eta^{}_{n - 1}, \eta^{}_n). \label{sea}
\end{multline}

\subsection{Full rqKZ equation}

\begin{figure}[ht]
\centering
\includegraphics{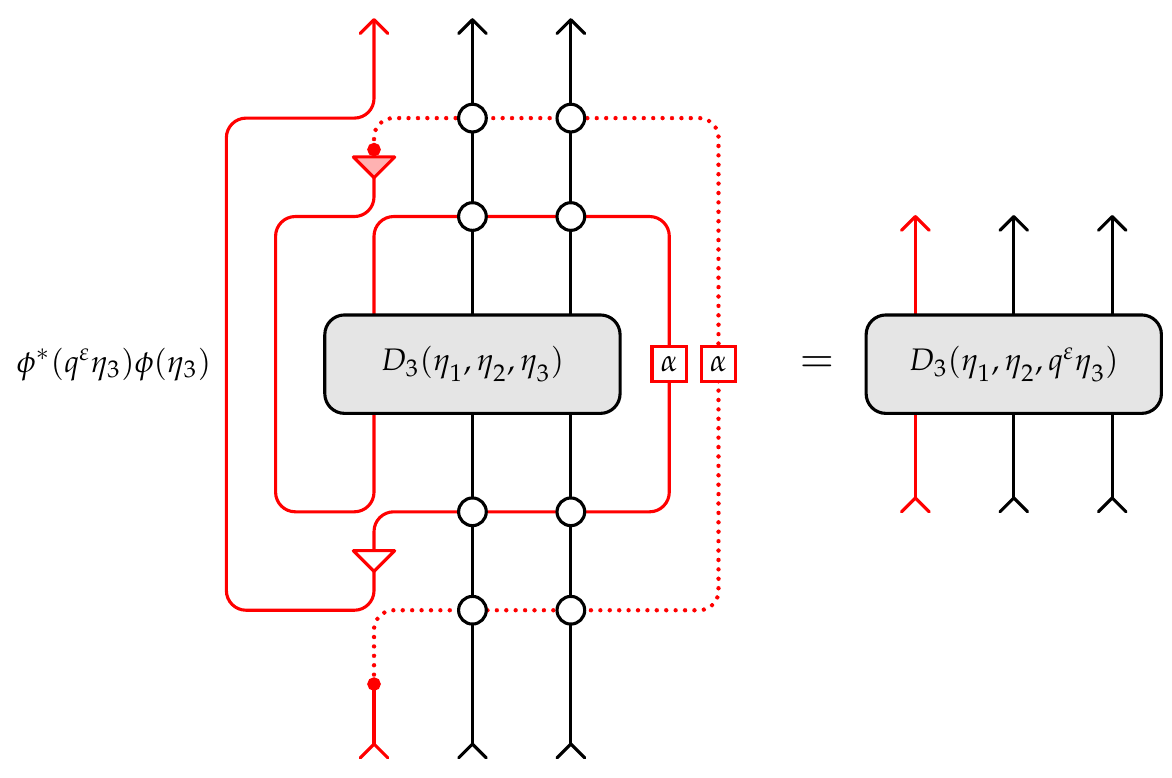}
\caption{}
\label{f:dqkze}
\end{figure}

Combining equations (\ref{andn}) and (\ref{bndn}), we come to the final reduced qKZ equation
\begin{multline*}
\phi^*(q^\lambda \eta_n) \phi(\eta_n) \, B_n(\eta_1, \ldots, \eta_{n - 1}, q^\lambda \eta_n) \big( A_n(\eta_1, \ldots, \eta_{n - 1}, \eta_n) (D_n(\eta_1, \ldots, \eta_{n -1}, \eta_n)) \big) \\
{} = D_n(\eta_1, \ldots, \eta_{n -1}, q^\lambda \eta_n).
\end{multline*}
The graphical image of this equation can be obtained by combining the graphical equations given in figures \ref{f:fe} and \ref{f:se}, see figure \ref{f:dqkze}.
Now we have two additional spaces, $V$ and $V^*$, and numerate the spaces as $0', 0, 1, \ldots, n$.  Combining equations (\ref{fea}) and (\ref{sea}), we obtain the following full reduced qKZ equation
\begin{align*}
& \phi(\eta_n) \phi^*(q^\lambda \eta_n) \, \tr_{0'} \tr_0 \big( A_{V^*}^{(0)}(\alpha) \check R^{(01)}_{V | V^*}(\eta_1 | q^\lambda \eta_n) \ldots R^{(n - 2, n - 1)}_{V | V^*}(\eta_{n - 1} | q^\lambda \eta_n) \\* 
& \hspace{11.8em} A_V^{(0')}(\alpha) \check R^{(0'0)}_{V | V}(\eta_1 | \eta_n) \ldots \check R^{(n - 3, n - 2)}_{V | V}(\eta_{n - 1} | \eta_n) \\*
& \hspace{13.5em} D^{(0', 0, 1, \ldots, n - 2)}_n(\eta^{}_1, \ldots \eta^{}_{n - 1}, \eta^{}_n) \\*
& \hspace{11.8em} \mathbbm{X}^{(n - 2, n - 1))}_V \check R^{(n - 3, n - 2)}_{V | V}(\eta_n | \eta_{n - 1}) \ldots \check R^{(0' 0)}_{V | V}(\eta_n | \eta_1) \\*
& \hspace{10.5em} \mathbbm{1}_{V^*}^{(n - 1, n)} \check R^{(n - 2, n - 1)}_{V^* | V}(q^\lambda \eta_n | \eta_{n - 1}) \ldots \check R^{(0 1)}_{V^* | V}(q^\lambda \eta_n | \eta_1) \big) \\*
& \hspace{25.6em} {} = D_n(\eta^{}_1, \ldots \eta^{}_{n - 1}, q^\lambda \eta^{}_n). 
\end{align*}
This is the main result of the present paper. In fact, we have the equation satisfied by the zero temperature correlation functions of the chain associated with the loop Lie algebra $\uqlg$. To investigate correlation function at arbitrary temperature one should return to equations (\ref{andnf}) and (\ref{bndnf}).

\section{Conclusions}

We have derived the reduced qKZ equation for the quantum integrable system related
to an arbitrary quantum loop algebra. The main feature of the general case compared to the simplest $\mathfrak{sl}_2$-case is that the first fundamental representation does not coincide with its dual, and so, to obtain a closed form of the reduced qKZ equation, two successive steps are needed. We have demonstrated that all necessary
unitarity and crossing relations follow from the properties of the algebra. The status of the initial condition is not completely clear. From one side, we do not see how it can be obtained from the properties of the algebra. From the other side, as we know, all $R$-operators found in the framework of the quantum group approach satisfy this condition.

Our result refers to the zero temperature case. In fact, intermediate equations (\ref{andnf}) and (\ref{bndnf}) can be used as a starting point to investigate the nonzero temperature case. The corresponding consideration for the quantum loop algebra $\uqlslii$ can be found in paper \cite{AufKlu12}. It should be noted that in the general case some additional problems arise. We hope to return to this later.

\section*{Acknowledgments}

We thank our colleagues and coauthors H.~Boos and F.~G\"ohmann for numerous fruitful discussions. KhSN was supported in part by the DFG grant \# BO3401/31 and by the 
Russian Academic Excellence Project `5-100'. AVR thanks the Max Plank Institute for Mathematics in Bonn for the warm hospitality extended to him during the work on this paper. This work was also supported in part by the RFBR grant \# 20-51-12005.

\appendix

\stepcounter{section}

\section*{Appendix: Graphical derivation of rqKZ equation}

\subsection{First equation}

\begin{figure}[!p]
\begin{minipage}[c][\textheight]{\textwidth}
\centering
\includegraphics [scale = .85] {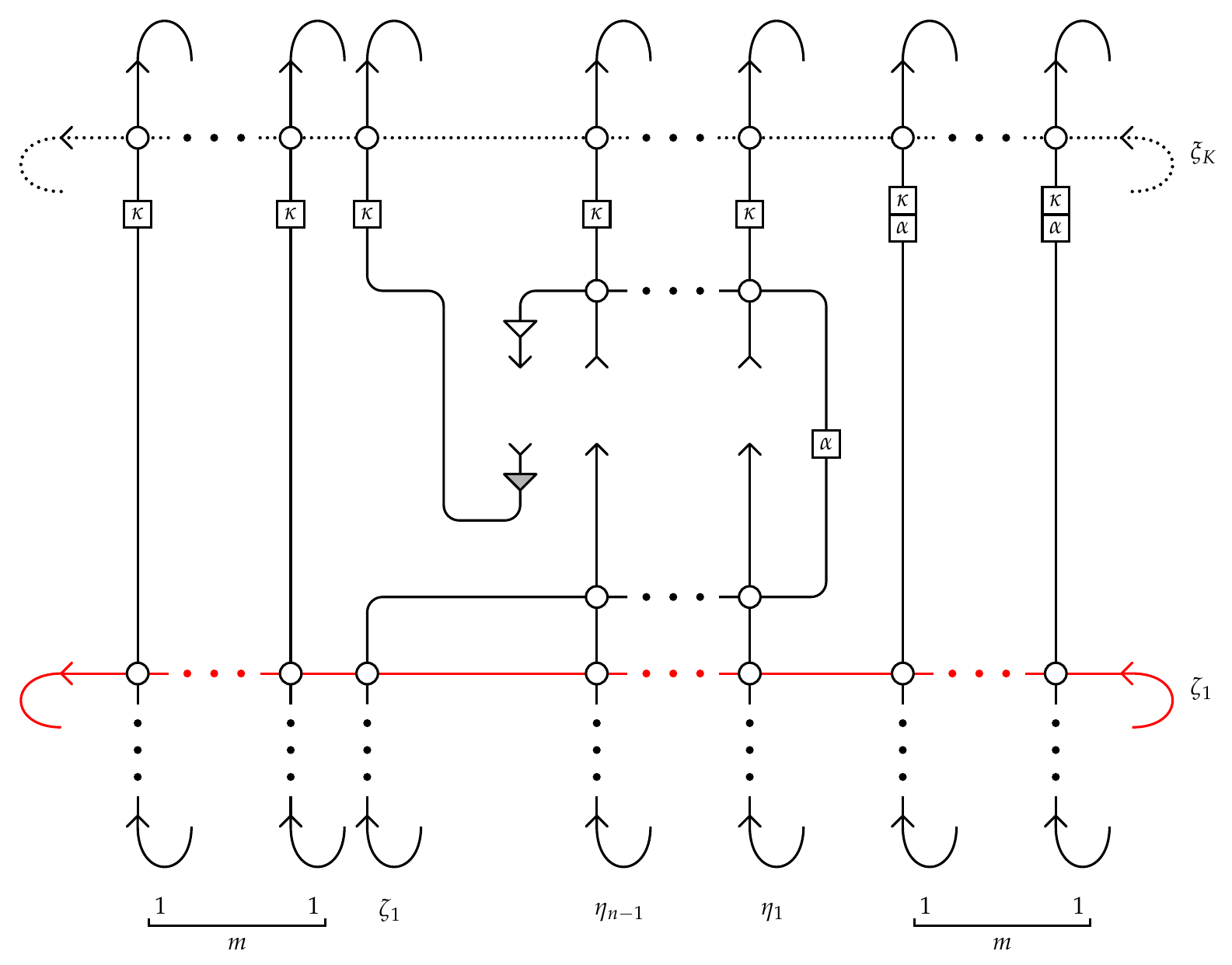}
\caption{}
\label{f:drqkza}
\vspace{1em}
\centering
\includegraphics [scale = .85] {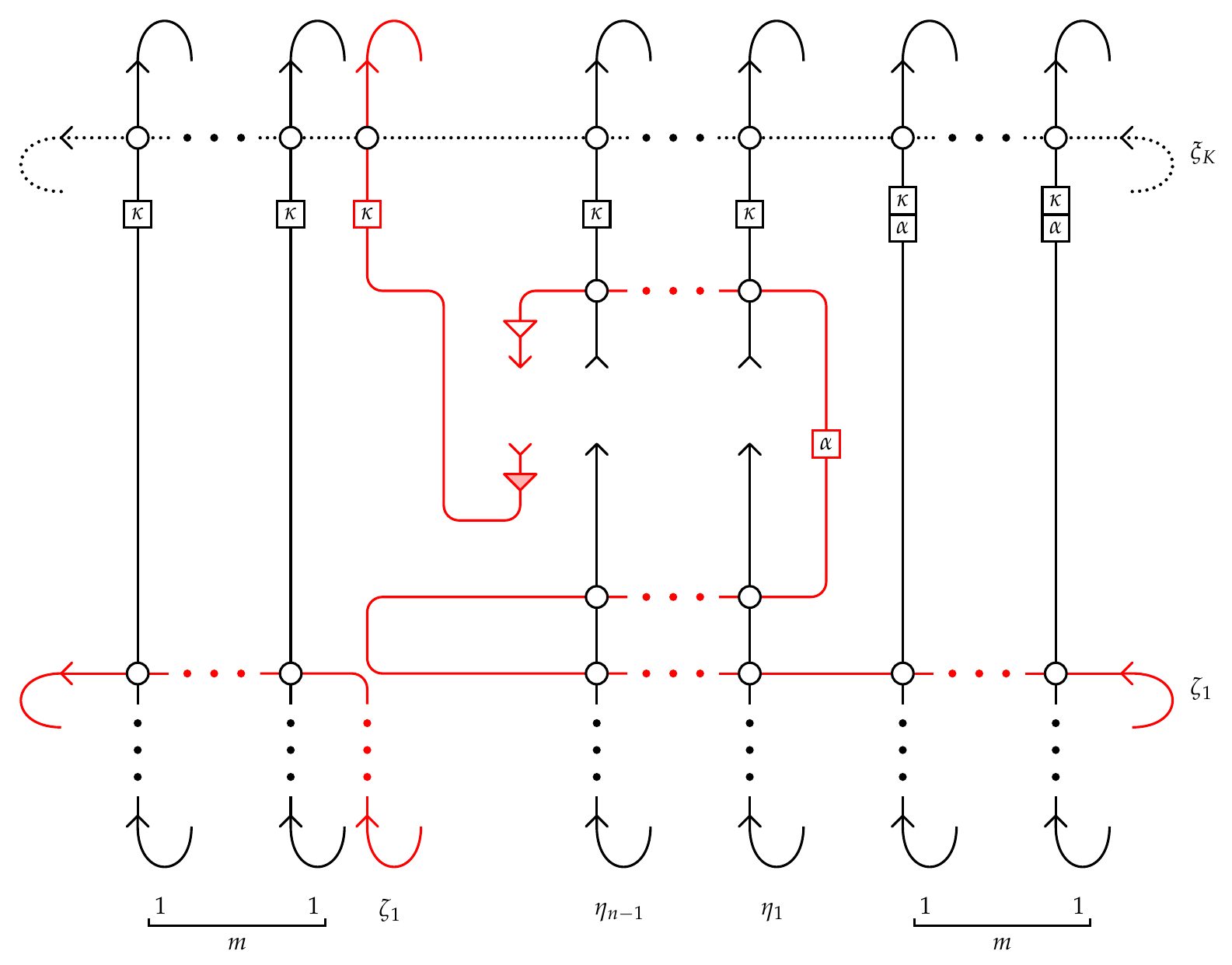}
\caption{}
\label{f:drqkzb}
\end{minipage}
\end{figure}

The initial configuration for the graphical derivation of the first part of
the reduced qKZ equation is given in figure \ref{f:drqkza}. The figure
represents the action of an operator, which we denote by $A_n(\eta_1, \ldots,
\eta_n)$, on $Z_{n, m, N}(\eta_1, \ldots, \eta_n) D_{n, m, N}(\eta_1, \ldots,
\eta_n)$. We mark by red the lines with the spectral parameter $\zeta_1$. The
triangle and the filled triangle corresponding to the operator $X_V$ and its
inverse are introduced to use further the crossing relations depicted in
figures \ref{f:crcni} and \ref{f:crcnii}. We put $\eta_n = \zeta_1$, use the
initial condition \ref{f:ici} and proceed to figure \ref{f:drqkzb}.

\begin{figure}[!p]
\begin{minipage}[c][\textheight]{\textwidth}
\centering
\includegraphics [scale = .85] {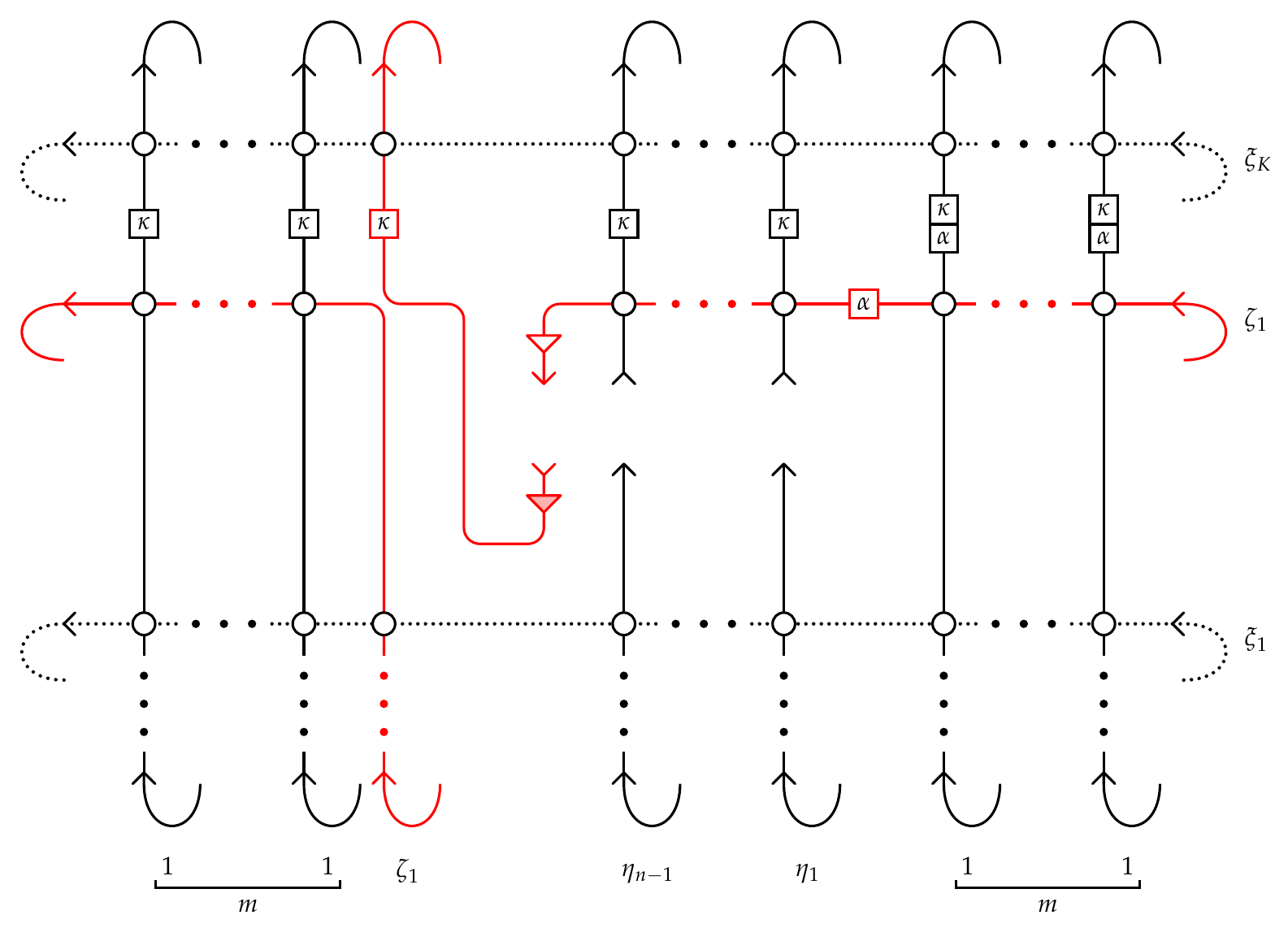}
\caption{}
\label{f:drqkzc}
\vspace{1em}
\centering
\includegraphics [scale = .85] {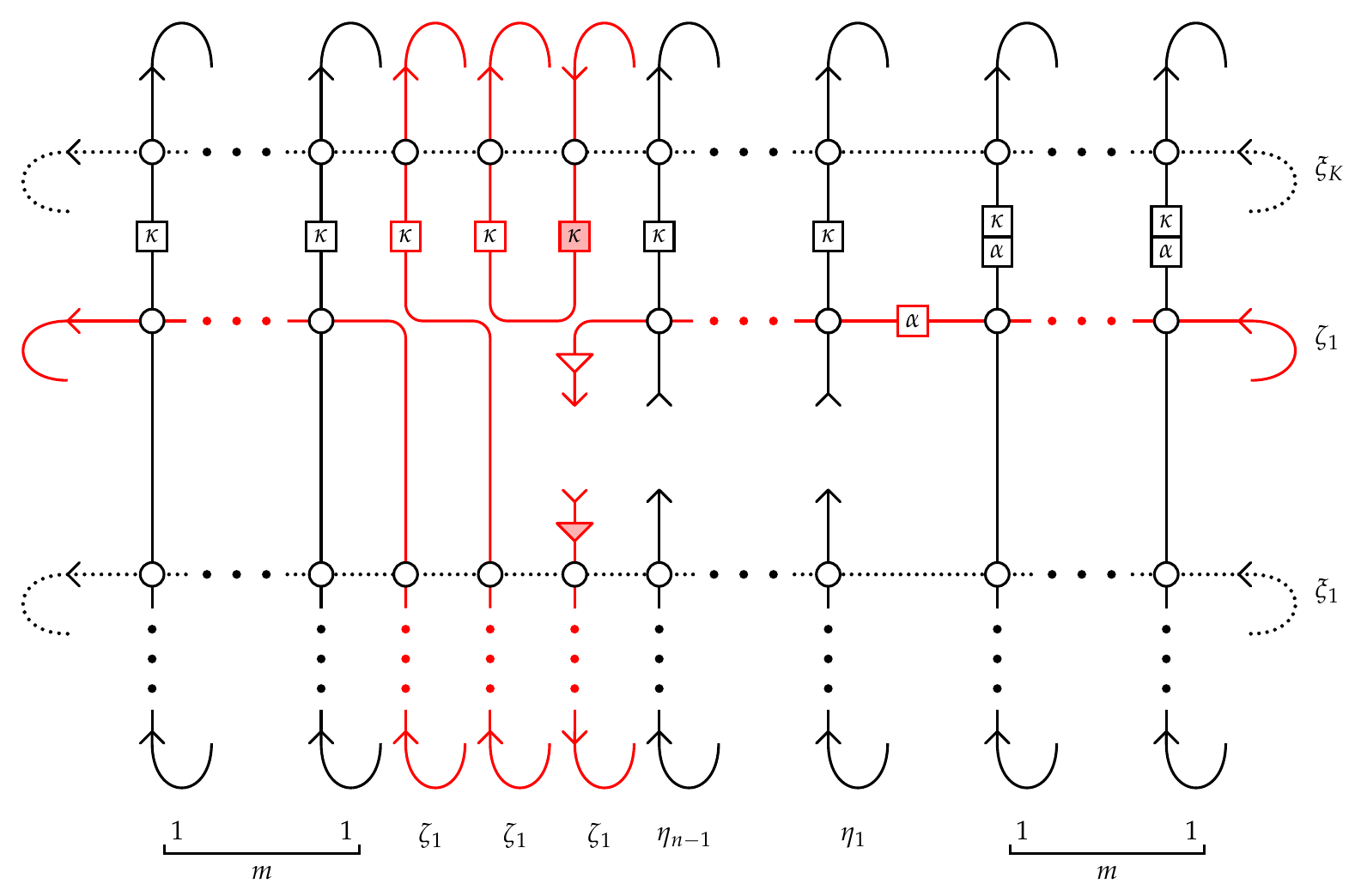}
\caption{}
\label{f:drqkzd}
\end{minipage}
\end{figure}

We pull off the emerging loop and raise the arising corner of the red line to the free corner above. This leads us to figure \ref{f:drqkzc}.

We move the `swing seat' down, then back up and front down again. To pass
through horizontal lines we use the unitarity relations \ref{f:urani} and
\ref{f:urcni}. After that we insert the product $A^{-1}_V(\kappa)
A^{}_V(\kappa)$ into the `swing seat' and go to figure \ref{f:drqkzd}.

\begin{figure}[!p]
\begin{minipage}[c][\textheight]{\textwidth}
\centering
\includegraphics [scale = .85] {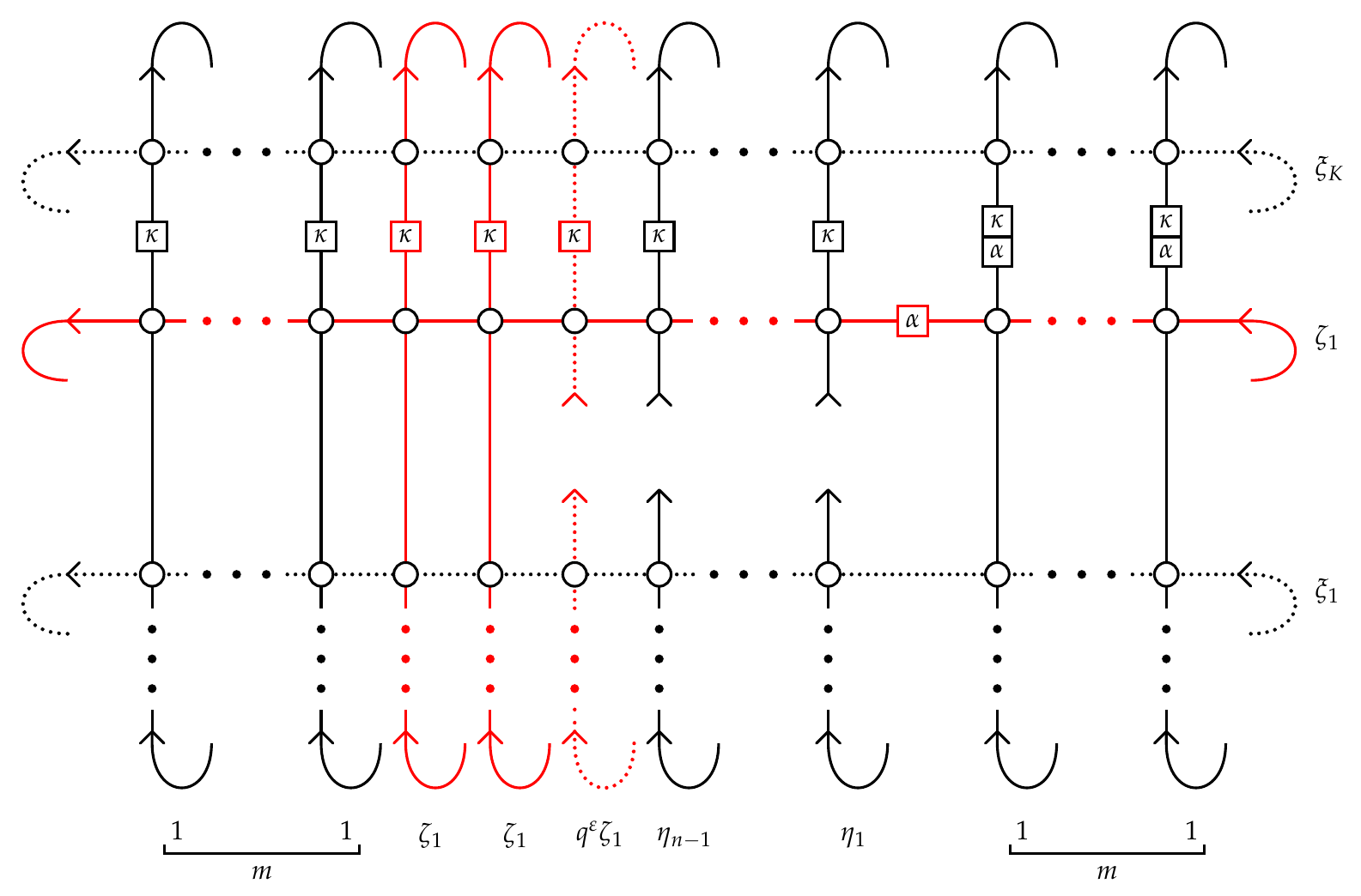}
\caption{}
\label{f:drqkze}
\vspace{1em}
\centering
\includegraphics [scale = .85] {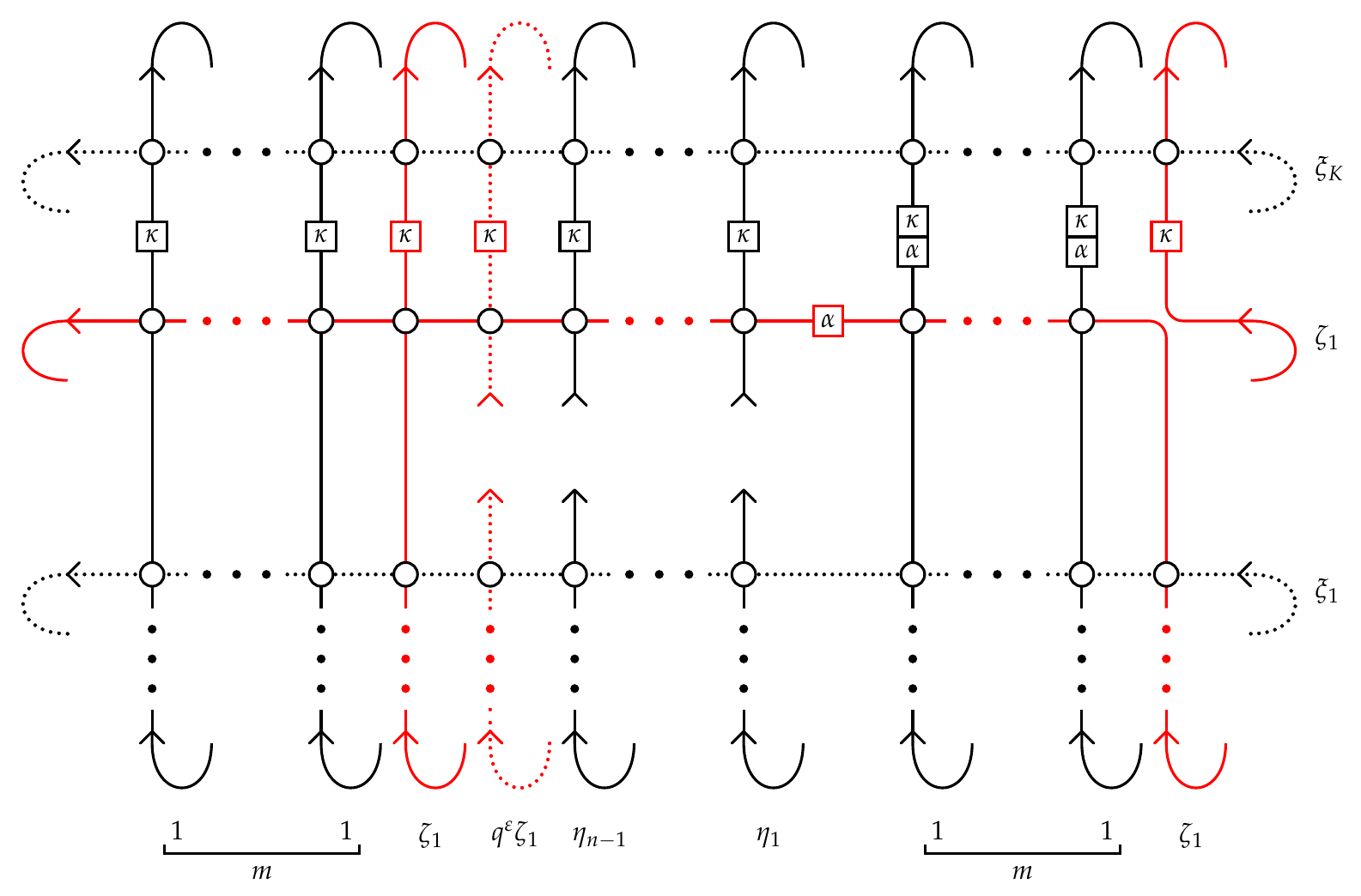}
\caption{}
\label{f:drqkzf}
\end{minipage}
\end{figure}

Now we restore all split vertices and, using the equations represented by figures \ref{f:crcni}, \ref{f:crcnii} and \ref{f:aca}, and the unitarity relations given in figures \ref{f:urcni} and \ref{f:urbnii}, reverse the direction of the red vertical line which goes from top to bottom. The commutativity of the operators $X_V$ and $A_V(\nu)$ is also used. We acquire the overall factor $\prod_{i = 1}^N D(q^\lambda \zeta_1, \zeta_i) D(q^\lambda \zeta_1, \xi_i)^{-1}$, where $\lambda = (\theta | \theta) \svee{h}/s$ and $D(\zeta, \eta)$ is defined by equation~(\ref{dze}). We keep this factor in mind. It is clear that the arising red dotted line is associated with the spectral parameter $q^\lambda \zeta_1$. After all that we come to figure \ref{f:drqkze}.

We move the leftmost red line behind the scene to the rightmost position, use the initial condition \ref{f:ici} and obtain the configuration given in figure \ref{f:drqkzf}.

The next task is to find the right place for the red box with the label $\alpha$. We use iteratively the graphical equation given in figure \ref{f:roaom} and proceed to the next figure.

\begin{figure}[!p]
\begin{minipage}[c][\textheight]{\textwidth}
\centering
\includegraphics [scale = .85] {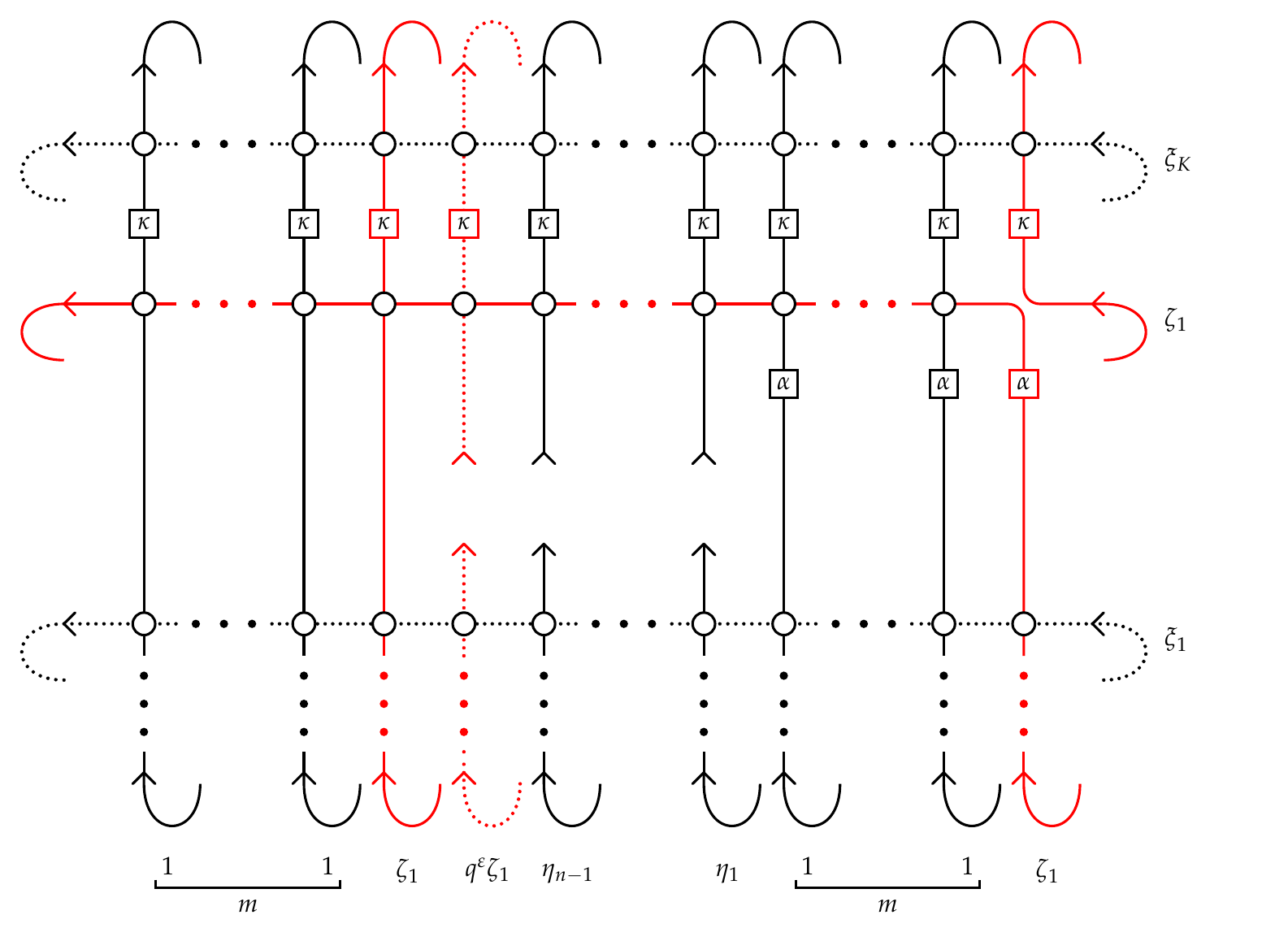}
\caption{}
\label{f:drqkzg}
\vspace{1em}
\centering
\includegraphics [scale = .85] {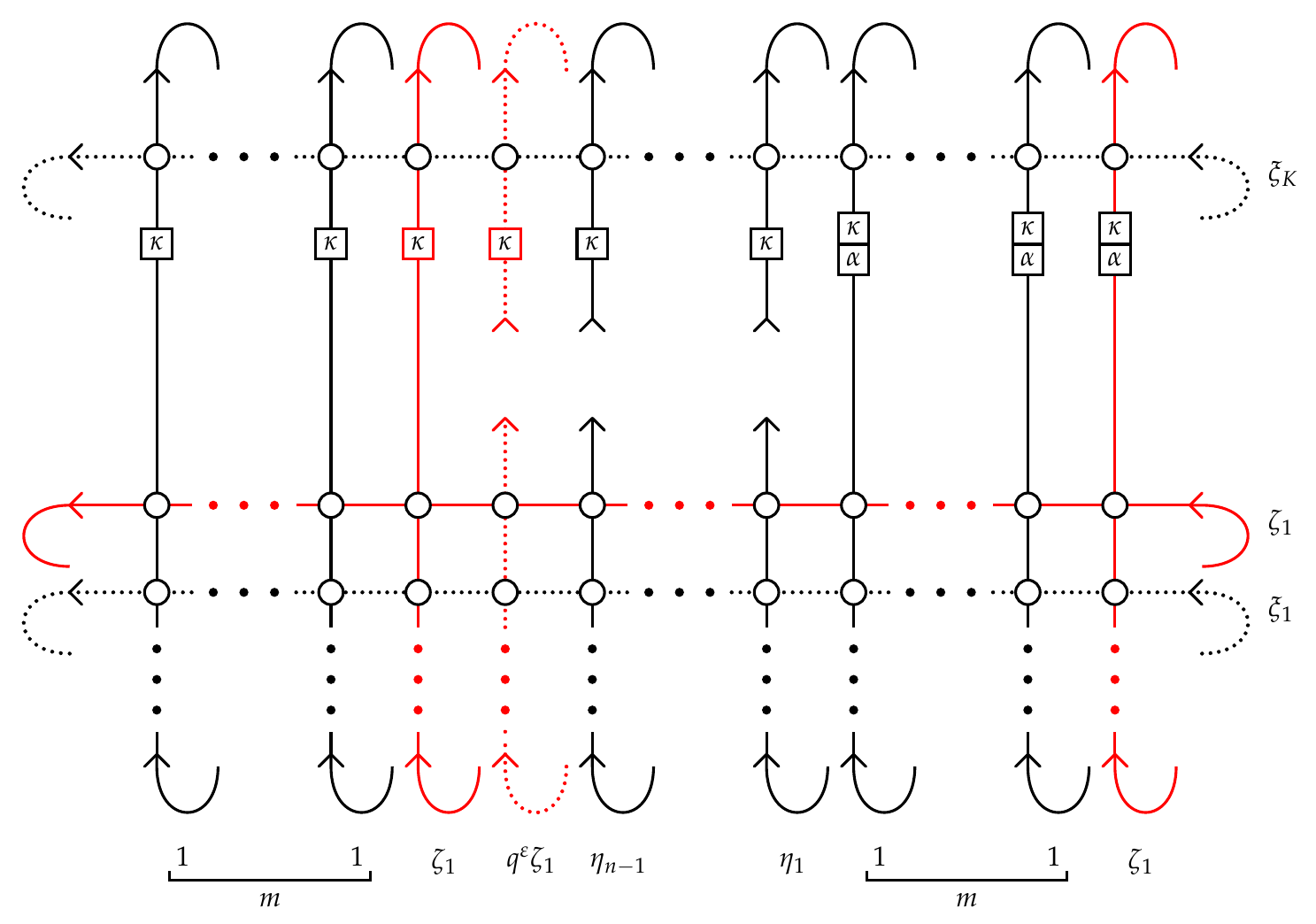}
\caption{}
\label{f:drqkzi}
\end{minipage}
\end{figure}

The use of the commutativity equation \ref{f:ctoa} allows us to order all twists. The last step is pretty cosmetic. We move the horizontal line with the spectral parameter $\zeta_1$ to the position where it was at the very beginning and stop at figure \ref{f:drqkzi}.

\subsection{Second equation}

\begin{figure}[!p]
\begin{minipage}[c][\textheight]{\textwidth}
\centering
\includegraphics [scale = .85] {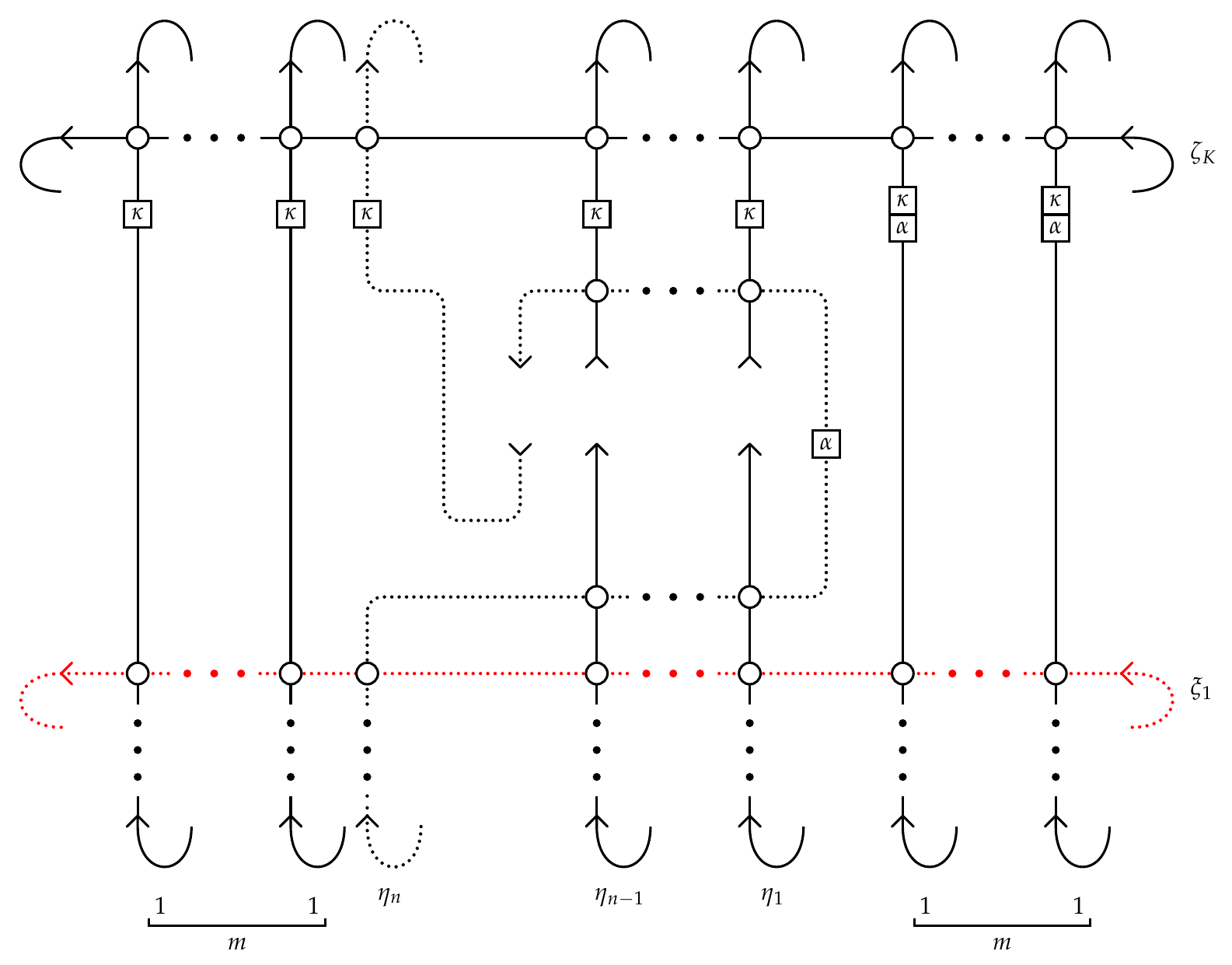}
\caption{}
\label{f:drqkzj}
\vspace{1em}
\centering
\includegraphics [scale = .85] {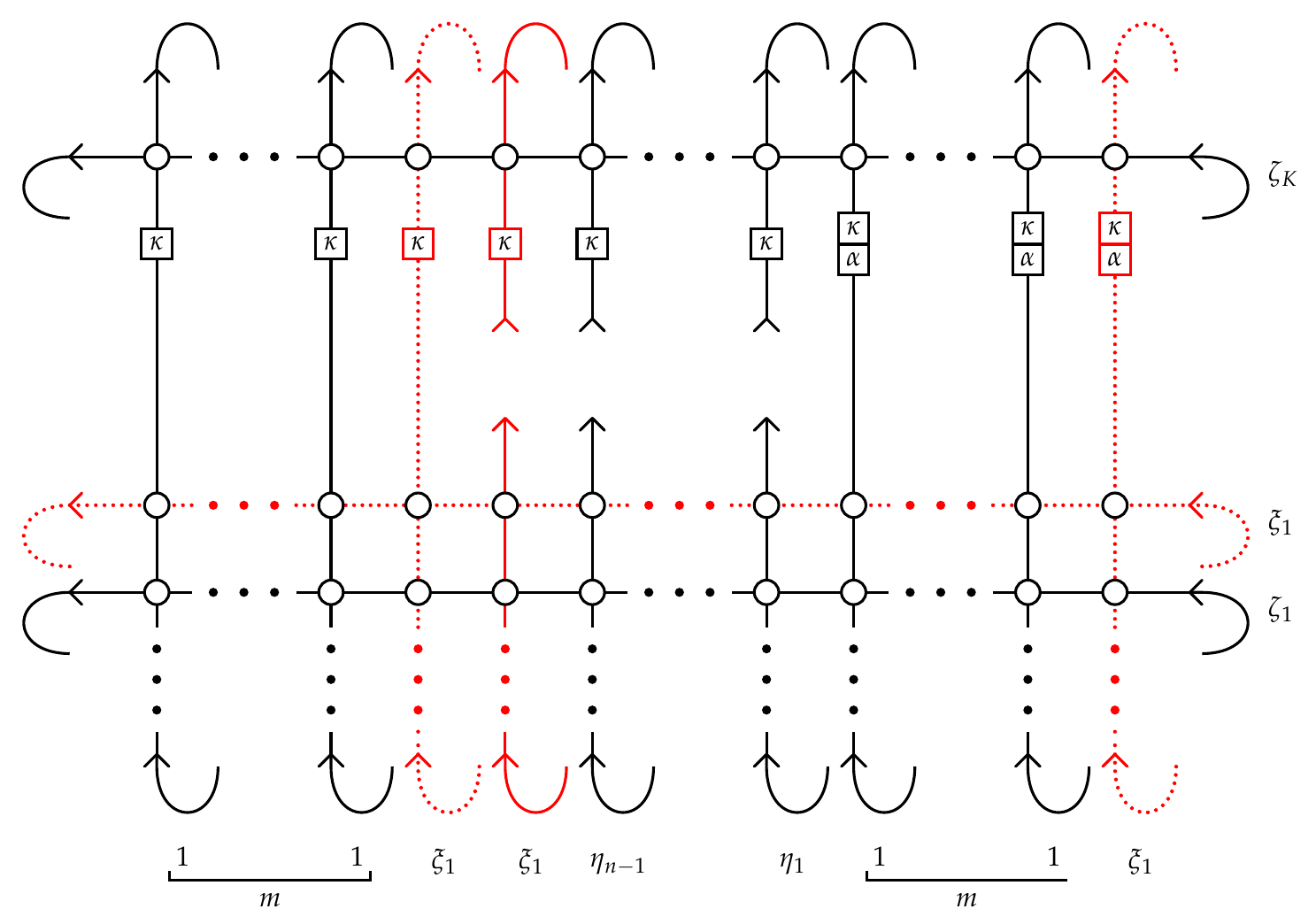}
\caption{}
\label{f:drqkzr}
\end{minipage}
\end{figure}

The initial point of the graphical derivation of the second part of the reduced qKZ equation is given in figure \ref{f:drqkzj}. Note that to apply the corresponding initial condition we make some rearrangement of the horizontal transfer operators using their commutativity.  The figure represents the action of an operator, which we denote by $B_n(\eta_1, \ldots, \eta_n)$, on the operator $Z_{n, m, N}(\eta_1, \ldots, \eta_{n - 1}, \eta^*_n)D_{n, m, N}(\eta_1, \ldots, \eta_{n - 1}, \eta^*_n)$. Now we mark by red the lines with the spectral parameter $\xi_1$. We put $\eta_n = \xi_1$ and perform transformations similar to those which we made in the derivation of the first part.

The final point of the graphical derivation of the second part of the reduced qKZ equation can be seen in figure \ref{f:drqkzr}.

\end{document}